\documentclass[12pt]{article}
\pdfoutput=1 

\usepackage[round]{natbib}

\usepackage{rotating}
\usepackage{graphicx}

\usepackage{amsthm,amsmath} 
\usepackage{amssymb}
\usepackage{subcaption}
\begin{document}

\thispagestyle{empty}
\baselineskip=28pt

\begin{center}
	{\LARGE{\bf Joint Hierarchical Gaussian Process Model with Application to Forecast in Medical Monitoring }}
\end{center}

\baselineskip=12pt

\vskip 2mm
\begin{center}
	Leo L. Duan
	\footnote{\baselineskip=10pt Department of Mathematical Sciences, University of Cincinnati},
	John P. Clancy
	\footnote{\baselineskip=10pt Division of Pulmonary Medicine, Cincinnati Children's Hospital Medical Center}
	and
	Rhonda D. Szczesniak
	\footnote{\baselineskip=10pt Division of Biostatistics and Epidemiology, Cincinnati Children's Hospital Medical Center}
	\footnote{\baselineskip=10pt Corresponding author. Address: 3333 Burnet Ave, MLC 5041, Cincinnati, OH 45229. Phone:(513)803-0563, email: rhonda.szczesniak@cchmc.org}
\end{center}

\baselineskip=12pt
A novel extrapolation method is proposed for longitudinal forecasting. A hierarchical Gaussian process model is used to combine nonlinear population change and individual memory of the past to make prediction. The prediction error is minimized through the hierarchical design. The method is further extended to joint modeling of continuous measurements and survival events. The baseline hazard, covariate and joint effects are conveniently modeled in this hierarchical structure. The estimation and inference are implemented in fully Bayesian framework using the objective and shrinkage priors. In simulation studies, this model shows robustness in latent estimation, correlation detection and high accuracy in forecasting. The model is illustrated with medical monitoring data from cystic fibrosis (CF) patients. Estimation and forecasts are obtained in the measurement of lung function and records of acute respiratory events.

\baselineskip=12pt
\par\vfill\noindent
{\bf KEY WORDS:}
{Extrapolation}, {Joint Model}, {Longitudinal Model}, {Hierarchical Gaussian Process}, {Cystic Fibrosis}, {Medical Monitoring}
\par\medskip\noindent
\clearpage\pagebreak\newpage
\pagenumbering{arabic}

\section{Introduction}

Forecasting for stochastic processes is commonly needed in geology, finance and clinical research; however, it is widely known that extrapolation is difficult and risky. Since the knowledge is limited to only the observed domain, without theoretical evidence, researchers tend to use simple functions for extrapolation.
Except for its conservativeness, this practice is often unrealistic and  overlooks many intrinsic properties, such as nonlinearity and stochastic fluctuations. This problem has been ameliorated by development in two fields of studies: time series and longitudinal data analysis. In the former, the variation in the past helps prediction in the future by the recursive relations. In the latter, the trend shared in batch data provides a reasonable guess for a given individual. Therefore, it is desirable to find a method that unifies these two fields.

In estimating the time-varying process behind the noise, nonparametric approaches such as penalized B-splines \citep*{eilers1996flexible} have been quite successful. However, one of the limitations is the subjective allocation of knots. For individual traces with only a few recorded points, it is still difficult to avoid over-fitting. Although methods like Bayesian Adaptive Regression Splines (BARS)  \citep*{dimatteo2001bayesian} have been proposed to address such issue, it is not feasible to use it for extrapolation.

An alternative approach is Gaussian process regression (GPR) \citep*{rasmussen2006gaussian}. By using a small number of (quite often as low as 1) parameters and a smooth covariance function, GPR avoids the use of knots and keeps the dimension fixed. This enables a fast estimation without sacrifices in robustness. In Eqn~\ref{GP}, assume $Y$ is a stochastic realization of time dependent function $  f( \boldsymbol t)$, with ${  \boldsymbol \epsilon} \sim {N}({ \boldsymbol 0}, \sigma_y^2 {  \boldsymbol I})$. If $   f( \boldsymbol t)$ is a Gaussian process and $ \boldsymbol \Sigma$ is formed by a covariance function that is differentiable with respect to time increments, then the posterior mean will also be a differentiable (smooth) function:

\begin{equation}
\begin{aligned}
&{ \boldsymbol{Y}}={ f( \boldsymbol t)}+ {\boldsymbol\epsilon}\\
&{  f( \boldsymbol t)}\sim {GP}({\boldsymbol\mu},{\boldsymbol \Sigma})\\
&{  f( \boldsymbol t)|\boldsymbol Y} \sim {N}(\boldsymbol\mu^*,\boldsymbol\Sigma^* )\\
&\boldsymbol\mu^*={{\boldsymbol\mu}+{ \boldsymbol\Sigma  (\boldsymbol\Sigma+}\sigma_y^2 {\boldsymbol I})^{-1}{({\boldsymbol Y}-{\boldsymbol \mu})}}\\
&\boldsymbol\Sigma^*={\boldsymbol \Sigma-\boldsymbol\Sigma (\boldsymbol\Sigma}+\sigma_y^2{\boldsymbol I})^{-1} {\boldsymbol\Sigma}
\end{aligned}
\label{GP}
\end{equation}

Major progress has been made in its predicting ability. For example, the Kriging estimator has been shown to be the best linear unbiased predictor (BLUP) and has been successful as a tool for interpolation\citep*{cressie1988spatial}. In Eqn~\ref{Kriging}, we use ${ \boldsymbol t}$ to denote the time vector of observed data, ${ \boldsymbol s}$ to denote the vector of prediction time and ${ \boldsymbol K( \boldsymbol s, \boldsymbol t)}$ to denote the their covariance:

\begin{equation}
\begin{aligned}
&{ f( \boldsymbol s)| \boldsymbol Y( \boldsymbol t)} \sim {N}(\boldsymbol\mu^* ,\boldsymbol \Sigma^* )\\
& \boldsymbol\mu^*={\boldsymbol\mu(\boldsymbol s)}+{\boldsymbol K(\boldsymbol s,\boldsymbol t)  (\boldsymbol\Sigma(\boldsymbol t)+}\sigma_y^2 { \boldsymbol I})^{-1}{({\boldsymbol Y(\boldsymbol t)}-{\boldsymbol \mu(\boldsymbol t)})}\\
& \boldsymbol \Sigma^*= {\boldsymbol \Sigma(\boldsymbol s)-\boldsymbol K(\boldsymbol s,\boldsymbol t) ( \boldsymbol \Sigma(\boldsymbol t)}+\sigma_y^2{\boldsymbol I})^{-1} {\boldsymbol K'(\boldsymbol s,\boldsymbol t)}
\end{aligned}
\label{Kriging}
\end{equation}

The magnitude of ${\boldsymbol K(\boldsymbol s,\boldsymbol t)}$ is usually reversely dependent on the distance measure $|s_i-t_j|$. In interpolation, these distances remain moderate since ${s}$ is inside the domain of ${t}$; whereas in extrapolation, all $|s_i-t_j|$ increase monotonically. As a result, the prediction mean monotonically reduces to ${\boldsymbol \mu(\boldsymbol s)}$ and prediction variance increases to ${\boldsymbol \Sigma(\boldsymbol s)}$. Therefore, slowing down the reduction of ${\boldsymbol K(\boldsymbol s,\boldsymbol t)}$ and improving the estimate of ${\boldsymbol \mu(\boldsymbol s)}$ are essential to achieve reasonable forecasting results. The Gaussian process functional regression (GPFR) model \citep*{shi2007gaussian} was proposed to solve this problem. Similar to longitudinal data, the data described by Shi and colleagues are collected in batches. In the first step, B-splines are used to estimate the batch mean at ${t}$ and ${s}$; in the second step, this mean is used as ${\mu(t)}$ and ${\mu(s)}$ for individual extrapolation. This approach greatly improves the forecasting ability of Gaussian process.

On the other hand, if the longitudinal data are collected at the same time as a related survival event, a joint model is commonly adopted for improved estimation and inference. Longstanding motivation for methods to link longitudinal and time-to-event data originated from human immunodeficiency virus (HIV)\citep*{song2002semiparametric}. Most recent developments of joint longitudinal-survival models have been accompanied by online calculators for the purposes of real-time individual prediction of prostate cancer recurrence  \citep*{taylor2013real}. Nevertheless, some challenges remain in the field of joint modeling: the estimation is difficult in the baseline hazard and full likelihood; the forecasting is unstable, especially in recurrent survival event modeling; the association between two responses lacks a realistic interpretation. The survival function specified in the Cox relative risk model \citep*{cox1972regression} takes the form of  Eqn~\ref{Cox_survival}. 

\begin{equation}
\begin{aligned}
S(T \ge t_2|T>t_1)=exp\{-\int_{t_1}^{t_2} \lambda_0(u)exp\{\boldsymbol X\boldsymbol \beta+f(u)\}du\}
\end{aligned}
\label{Cox_survival}
\end{equation}

The use of $t_1$ is to accommodate the possibility of a recurrent event. In the case of nonrecurring events, we simply set $t_1=0$. Since the data are collected at discrete time points, approximation is usually needed to evaluate the integral. However, if one needs to forecast multiple time points corresponding to recurring events, $t_1$ is random and Eqn~\ref{Cox_survival} becomes intractable. To tackle these problems, we adopt the discrete Cox relative risk model provided in the same article \citep*{cox1972regression}. The estimation and prediction now have a tractable solution in closed form. Details are described later in section 2.2.

The major novelty in our approach is that we use two hierarchical Gaussian processes for both longitudinal and survival submodels. In the longitudinal part, the first hierarchy enables the sharing of the trajectory trend among subjects; and the second captures the individual deviations through a time-series covariance function. In the survival part, the first Gaussian process acts as a smoother for the baseline; the second Gaussian process serves as a time-varying frailty term. Using the of shared parameter framework \citep*{vonesh2006shared}, we set up the association between the two responses through a time-varying covariance. This hierarchical structure enables a reliable extrapolation by combining a nonlinear population trend, individual autocorrelation and joint effect. The model is straightforward and the estimation procedure is completely likelihood-driven and single-staged. The computation is demonstrated in a fully Bayesian framework and Expectation-Maximization algorithm.

The remainder of the article is organized as follows. In Section 2, we present details for the proposed hierarchical Gaussian process (HGP) model, its extension as a survival model and the joint hierarchical Gaussian process (JHGP) model. In Section 3, we present the simulation studies and assess forecasting performance. In Section 4, we apply the JHGP model to clinical data from patients with cystic fibrosis. Concluding remarks and discussion are presented in Section 5.

\section{Methods}

\subsection{Hierarchical Gaussian Process Model}

\subsubsection{Model Structure}
The records of  $\boldsymbol Y_{ij}$'s are assumed to be from a continuous stochastic process ${\boldsymbol Y_i(\boldsymbol t)}$ of subject $i$ at the $j$th time interval ($i=1,...,n$ and $j=1,...,n_i$). For simplicity of notation, we disregard other covariate effects for now and assume:

\begin{equation}
\begin{aligned}
&{\boldsymbol Y_i(\boldsymbol t)}={f_i(\boldsymbol t)}+\boldsymbol \epsilon      \text{ where }  \boldsymbol \epsilon \sim {N(0, \boldsymbol I \sigma_y^2)}\\
&{f_{i}(\boldsymbol t)|\boldsymbol \mu_y }\stackrel{indep.}{\sim}{ GP} [\boldsymbol \mu_y+\gamma_i \boldsymbol 1,{\boldsymbol V}_\psi\sigma_{\psi_i}^2] \text{ for } i=1,...,n\\
&{ \boldsymbol \mu_y} \sim {GP[ \boldsymbol 0,\boldsymbol V_{\mu_y}\sigma_{\mu_y}^2]}
\end{aligned}
\label{eqn_HGP}
\end{equation}

It is worth noting that there are multiple independent copies of ${f_i(\boldsymbol t)}$ but only one copy of $\boldsymbol \mu_y$. In other words, $\boldsymbol \mu_y$ is the shared mean process for all subjects. The time span of ${\boldsymbol \mu_y}$ is equal to the full span of the longitudinal data, where one of the ${f_i(\boldsymbol t)}$ functions is limited to the subject's first and last observation (or censoring) time. We add $\gamma_i$ to accommodate individual differences at the beginning of each trajectory. Each individual has a different scale parameter $\sigma^2_\psi$, but shares the same correlation matrix $\boldsymbol V_\psi$.

We choose a differentiable covariance function (with respect to $\Delta t=t_i-t_j$) to generate ${\boldsymbol V_{\mu_y}}$. One example of such a function is the squared exponential  $\{ exp(-\frac{(\Delta t)^2}{2 {\lambda}^2})\}_{i,j}$. This guarantees that $\boldsymbol \mu_y$ is a smooth function in time. The differentiability replaces the role of knots in spline-based approaches, thereby avoiding the dimension change problem. On the other hand, we choose a non-differentiable time-series function to generate ${\boldsymbol V_\psi}$. For example, we use an AR(1) covariance $\{ \rho^{|i-j|}\}_{i,j}$ ($-1 < \rho < 0$) to force $f_i(\boldsymbol t)$ to have a trajectory that resembles random walk.

\subsubsection{Predictive Distribution}

Forecasts $f_i(\boldsymbol s)$ at time vector ${\boldsymbol s}$  can then be obtained by conditioning on $\boldsymbol \mu_y$, ${\boldsymbol V}_\psi\sigma_{\psi_i}^2$ and ${\boldsymbol Y_i(\boldsymbol t)}$.

\begin{equation}
\begin{aligned}
&{ \boldsymbol f_i(\boldsymbol s)|\boldsymbol \mu_y, {\boldsymbol V}_\psi,\boldsymbol Y_i(\boldsymbol t)} \sim {N}(\boldsymbol\mu^*,\boldsymbol\Sigma^*)\\
&\boldsymbol\mu^*={\boldsymbol \mu_y(\boldsymbol s)}+\gamma_i {\boldsymbol 1_s}+{\boldsymbol  K(\boldsymbol s,\boldsymbol t)  ({\boldsymbol V}_\psi\sigma_{\psi_i}^2+}\sigma_y^2 { \boldsymbol I})^{-1}{({\boldsymbol Y(\boldsymbol t)}-{\boldsymbol \mu_y(\boldsymbol  t)-\gamma_i {\boldsymbol 1_t}})}, \\
 &\boldsymbol\Sigma^*={\boldsymbol \Sigma(\boldsymbol s)-\boldsymbol K(\boldsymbol s,\boldsymbol t) ({\boldsymbol V}_\psi\sigma_{\psi_i}^2}+\sigma_y^2{\boldsymbol I})^{-1} {\boldsymbol K'(\boldsymbol s,\boldsymbol t)})
\end{aligned}
\label{predictive_dist}
\end{equation}
which is similar to Eqn~\ref{Kriging}. The main difference in Eqn~\ref{predictive_dist} is that ${\boldsymbol \mu_y(\boldsymbol s)}$ and ${\boldsymbol \mu_y(\boldsymbol t)}$ are now subsets of $\boldsymbol \mu_y$, which is a Gaussian process instead of a simple function.

\begin{figure}[!H]

          \begin{subfigure}[b]{.3\columnwidth}
 	\centering\includegraphics[width=1\columnwidth]{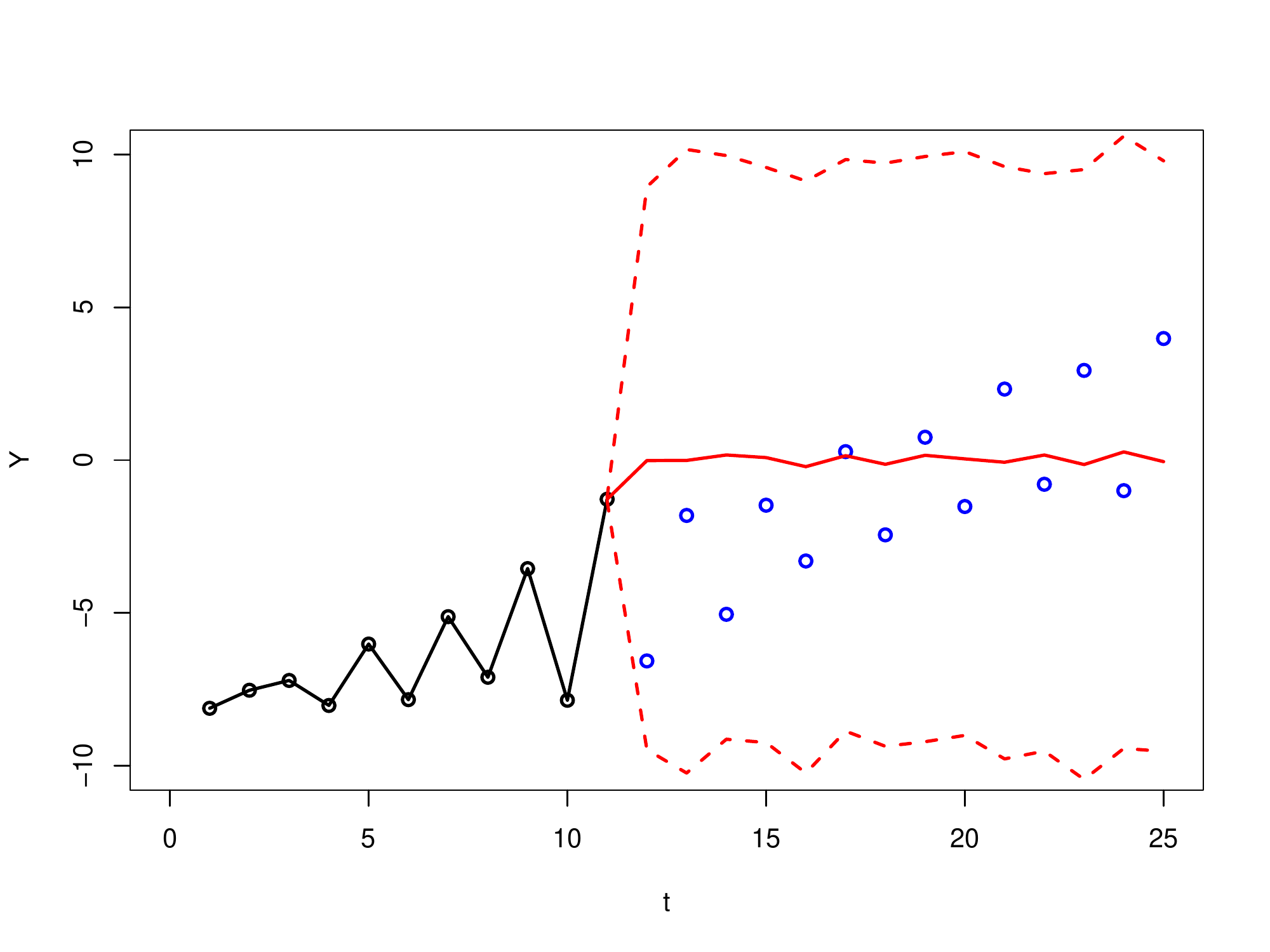}
	\subcaption{Prediction using the autocorrelation of one Gaussian process}
		\end{subfigure}
          \begin{subfigure}[b]{.3\columnwidth}
		\centering\includegraphics[width=1\columnwidth]{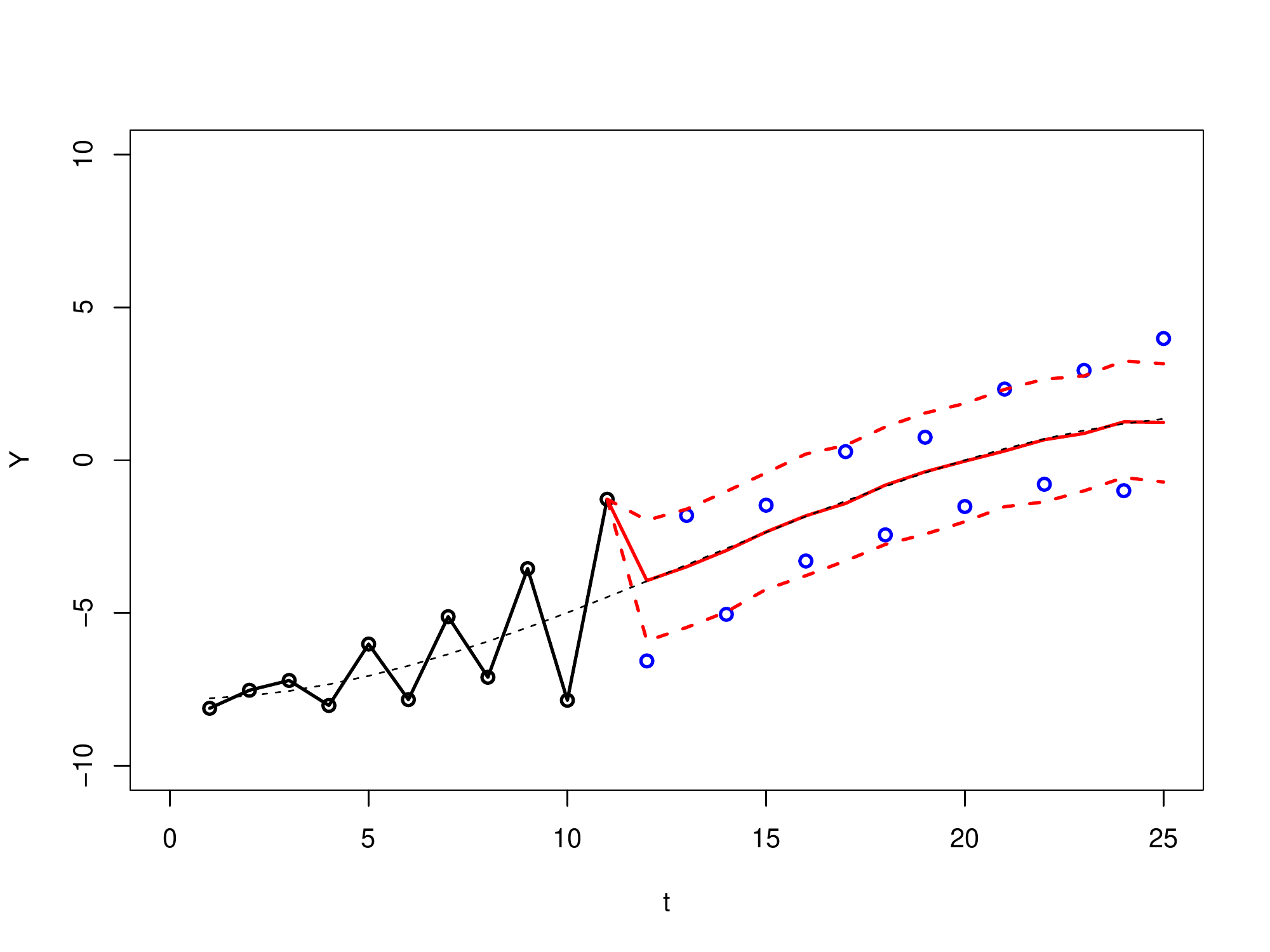}
		\subcaption{Prediction using the estimated $\mu$ from the other subjects}
		\end{subfigure}
        \begin{subfigure}[b]{.3\columnwidth}
		\centering	\includegraphics[width=1\columnwidth]{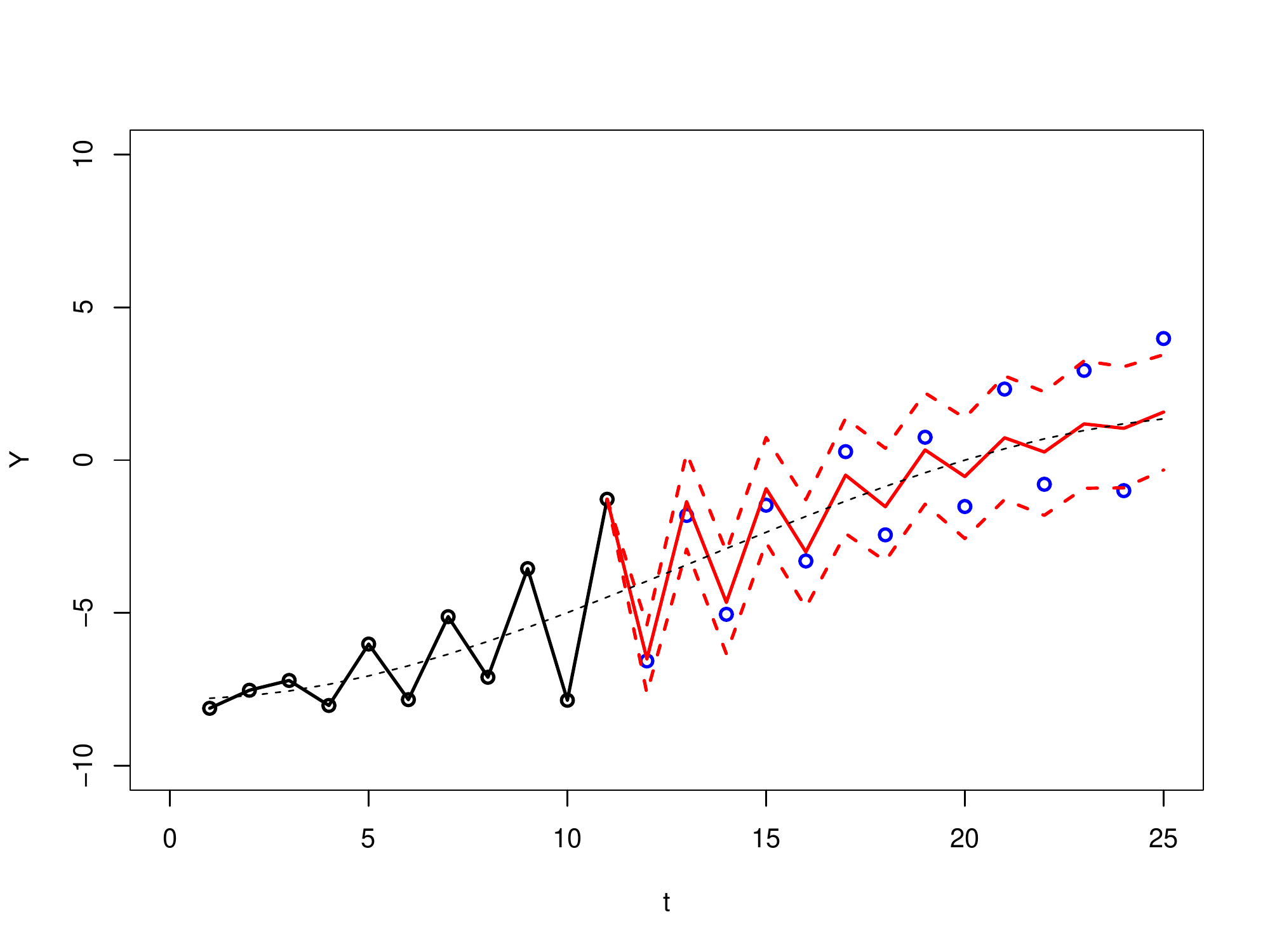}
			\subcaption{Extrapolation based on $\mu$ and the autocorrelation}
			\end{subfigure}
\caption{Forecasting using HGP model results in better mean estimates and regulated prediction errors, compared with using only one Gaussian process}
\label{forecast_sims}
\end{figure}

The benefits of having two Gaussian processes for prediction are illustrated in Figure~\ref{forecast_sims}. The test samples are first generated in batch ($n=50$), then we randomly select one subject and remove the corresponding second half of the observed points (shown in blue). We first fit each subject with an individual Gaussian process with the AR(1) covariance (Figure~\ref{forecast_sims}(a)). Although the fitted (black) line shows that the model has adequate flexibility, it cannot perform well in extrapolation: the red line rapidly reverts to the constant mean. This behavior is due to the small autocorrelation ($\rho$ close to $0$), caused by the heterogeneity of the observation. We next fit all subjects with a common Gaussian process $\mu$ with the squared exponential covariance, and examine its prediction performance for the individual with the masked data (Figure~\ref{forecast_sims}(b)). The prediction benefits from the similarity of trajectories among all the subjects. Since $ {\boldsymbol Y-\boldsymbol \mu_y}$ is more homogeneous than $\boldsymbol  Y$, we use the second individual Gaussian process (AR(1)) conditional on the estimate of $\boldsymbol \mu$  (Figure~\ref{forecast_sims}(c)). The magnitude of autocorrelation becomes larger ($\rho=-0.8$) and the decrease of $\boldsymbol  K(\boldsymbol s,\boldsymbol t)$ becomes slower. As a result, both the point estimates and credible intervals of the forecast greatly improve and become personalized.

Since an autoregressive covariance function is extensively used in our predictive distribution, it is worth mentioning that the prediction mean in Eqn~\ref{predictive_dist} is equivalent to the recursive forecast equation in time-series analysis. Denote $X_t$ as the discrete observation of ${Y(t)-\mu_y}$ at time point $t$:

\newtheorem{thm}{Definition}

\begin{thm}
If vector $\{X_t\}_t$ is from stationary AR(p) process, $$X_t=\sum_{i=1}^p \psi X_{t-i}+\epsilon_t$$\\
Then the pointer estimator for $X_{t+1}$:
$$\sum_{i=1}^{p}\psi_i X_{t+1-i}=
 \begin{bmatrix} \sigma^2_{1,t+1} & \cdots& \sigma^2_{t,t+1}\end{bmatrix}   
\begin{bmatrix} \sigma^2_{1,1} & \cdots& \sigma^2_{1,t}\\
 \vdots & \ddots& \vdots\\
\sigma^2_{t,1} & \cdots& \sigma^2_{t,t}\\
\end{bmatrix} ^{-1}
 \begin{bmatrix} X_1\\ \vdots\\X_{t}\end{bmatrix}$$
where $\{\sigma^2_{i,j}\}_{i,j}$ are the elements of covariance matrix of AR(p) process.
 \label{thm_1}
\end{thm}

The proof is left in Appendix. This equivalence holds for the multiple-step forecast by induction.

\subsection{Extension to the Survival Model}

\subsubsection{Model Structure}
The Cox relative risk model  \citep*{cox1972regression} has been widely used in survival analysis. We show the HGP can be adopted into the Cox model, thereby enabling forecasting with survival data.

As previously mentioned in Section 1, the discrete Cox model avoids the numerical integration, provides a closed-form solution to the baseline estimates and also the flexibility to incorporate recurrent events. The data set of discrete survival event is formed in the following procedures: for every event or censoring time $t$, assign it into discrete slot $k$; create the corresponding binary variable $R_k$ that takes value $1$ if event happens or $0$ if censored; use $R_0$ to denote the result of the first observation period; then fill all the periods between the $0$ and $k$ with $R_j=0$; in the case of a recurrent event, fill all the periods between two consecutive events with $R_j=0$. Let $\lambda_i(k)=\mathbb{P}(R_k=1|R_{\{j:l<j<k\}}=0)$, where $l$ is either the start time or the time of the last event if it is recurrent. The resulting $\lambda_i(k)$ is referred to as the discrete hazard function of individual $i$ at time $k$. The full likelihood of a survival event or censoring can be expressed as

\begin{equation}
\begin{aligned}
&\mathbb{P}(R_k=1,R_{\{j:l<j<k\}}=0)=\lambda_i(k)\prod_{j=l+1}^{k-1}\{1-\lambda_i(j)\}\\
&\mathbb{P}(R_k=0,R_{\{j:l<j<k\}}=0)=(1-\lambda_i(k))\prod_{j=l+1}^{k-1}\{1-\lambda_i(j)\}
\end{aligned}
\end{equation}
If we let $\lambda_0(k)$ denote the value of the baseline hazard at time $k$, then the discrete Cox relative risk model \citep*{cox1972regression} can be defined as:

\begin{equation}
\frac{\lambda_i(k)}{1-\lambda_i(k)}=\frac{\lambda_0(k)}{1-\lambda_0(k)}exp\{\boldsymbol  X_i\boldsymbol \beta+g_i(k)\}\end{equation}
where $X_i\beta$ represents the covariate effects and $g_i(k)$ is the time-dependent frailty.

By logarithm transformation, we have the equation:
\begin{equation*}
logit(\lambda_{i}(k))=logit(\lambda_{0}(k))+\boldsymbol X_i\boldsymbol \beta+g_i(k)
\end{equation*}

For ease of notation, we ommit $X_i\beta$ in the following equation. Since the baseline hazard $\lambda_{0}$ is commonly assumed to be continuous and smooth, the $logit$ link function is also a continuous and bijective function; therefore, the $logit(\lambda_{0})$ should also be a smooth function. Analogous to the two hierarchies in HGP, the baseline is a common smooth process shared by all subjects, the frailty $g_i$ is a subject-specific deviation. It is natural to use HGP to model these two processes.

For simplicity of notation, we use ${\boldsymbol H_{i}}=logit(\boldsymbol \lambda_{i})$ and $\boldsymbol \mu_H=logit(\boldsymbol \lambda_{0})$. The extended HGP model can be written as:

\begin{equation}
\begin{aligned}
&\boldsymbol \lambda_i=\frac{exp({\boldsymbol H_i})}{1+exp({\boldsymbol H_i})}\\
&{\boldsymbol H_i}|\boldsymbol \mu_{{h}}\stackrel{indep}{\sim} { GP} (\boldsymbol \mu_{{h}}+\eta_i {\boldsymbol 1}, { \boldsymbol V}_h \sigma^2_{h_i})\\
&\boldsymbol \mu_{{h}}{\sim} { GP} ({\boldsymbol 0}, {\boldsymbol V}_{\mu_{{h}}}{\sigma^2}_{\mu_{{h}}})
\end{aligned}
\label{eqn_HGP_logit}
\end{equation}

\noindent where $\eta_i$ provides an intercept shift relative to the baseline hazard, in order to accommodate diversity at the starting level. Note that the predictive distribution in Eqn~\ref{eqn_HGP_logit} is similar to Eqn~\ref{predictive_dist}.

\subsection{Joint Hierarchical Gaussian Process Model}
\subsubsection{Model Structure}
When continuous measurements and survival events are modeled jointly, the joint likelihood is factorized according to shared random parameter model  \citep*{vonesh2006shared}:
\begin{equation}
\mathbb{P}({\boldsymbol R,\boldsymbol Y})=\int \mathbb{P}({\boldsymbol R|\boldsymbol \psi})\mathbb{P}({\boldsymbol Y}|\boldsymbol \psi)\mathbb{P}(\boldsymbol \psi)d \boldsymbol \psi
\end{equation}
where ${\boldsymbol R}$ is the binary representation of the survival event, ${\boldsymbol Y}$ is the continuous response and $\boldsymbol \psi$ is their shared parameter. To enable time-dependency in ${\boldsymbol \psi}$, we assume $\boldsymbol \psi$ is the individual shared Gaussian process. The joint hierarchcial Gaussian process model is defined as:

\begin{equation}
\begin{aligned}
&{\boldsymbol \psi}_i\stackrel{indep.}{\sim} { GP} ({\boldsymbol 0}, { \boldsymbol V}_\psi\sigma^2_{\psi_i})\\
&\boldsymbol \mu_y \sim { GP}( \boldsymbol  0,\boldsymbol V_{\mu_y}\sigma^2_{\mu_y})\\
&\boldsymbol \mu_h \sim { GP} ( { \boldsymbol 0},\boldsymbol V_{\mu_h}\sigma^2_{\mu_h})\\
&{\boldsymbol Y}_i \sim { N}(  \gamma_i{\boldsymbol  1}+ \boldsymbol \mu_y+{\boldsymbol \psi}_i,  \boldsymbol I\sigma_y^2)\\
&{\boldsymbol H}_i=\eta_i{\boldsymbol  1}+\boldsymbol \mu_h+ {\boldsymbol  \psi}_i\phi\\
&{\boldsymbol R}_i \sim { Bin} (\frac{exp({\boldsymbol H}_i)}{1+exp({\boldsymbol H}_i)})\\
\end{aligned}
\end{equation}
The conditional distribution of ${\boldsymbol Y}_i$ and ${\boldsymbol H}_i$ is a multivariate Gaussian distribution:

\begin{equation}
\begin{aligned}
\begin{bmatrix}{\boldsymbol Y}_i\\{\boldsymbol H}_i\end{bmatrix}\biggr\rvert ( \gamma_i, \eta_i)\sim { N}
 \bigg\{
 \begin{bmatrix}\gamma_i{ \boldsymbol 1}\\\eta_i{\boldsymbol  1} \end{bmatrix},
  \begin{bmatrix}\boldsymbol V_{ \mu_y}\sigma^2_{\mu_y}+{\boldsymbol  V}_\psi\sigma^2_{\psi_i} &\phi{
  \boldsymbol  V}_\psi\sigma^2_{\psi_i}\\\phi{ \boldsymbol V}_\psi\sigma^2_{\psi_i}&  \boldsymbol V_{\mu_h}\sigma^2_{\mu_h}+\phi^2{\boldsymbol  V}_\psi\sigma^2_{\psi_i}\end{bmatrix}
  \bigg\}
\end{aligned}
\label{jhgp_cond}
\end{equation}


\subsection{Data Augmentation and Prior Elicitation}

To facilitate the rate of convergence of Markov chain Monte Carlo, we use the data augmentation technique for the logistic distribution \citep*{polson2012bayesian}.
\begin{equation*}
\frac{exp({H_{ij}}R_{ij})}{1+exp({H_{ij}})} \propto \int_{0}^{\infty} exp(-0.5\omega_{ij}H^2_{ij}+(R_{ij}-0.5) H_{ij}
 ) f(\omega_{ij})d\omega_{ij}
\end{equation*}
where $f(\omega_{ij})$ is the density of Polya-Gamma distribution $PG(1,0)$. Its posterior  $\omega_{ij}|H_{ij}\sim PG(1,H_{ij})$. $H_{ij}|\omega_{ij}$ is Gaussian distribution.

We choose objective and weakly informative priors for Bayesian analysis. For the parameters in the two Gaussian processes for the means $\mu_y$, $\mu_h$, we use the Jeffreys priors:

\begin{equation*}
\begin{aligned}
&[\theta_{\mu_y},\sigma^2_{\mu_y}]\propto \{tr(\boldsymbol U_{\mu_y}^2)-\frac{1}{n_t}tr(\boldsymbol U_{\mu_y})^2\}^{1/2} \frac{1}{\sigma^2_{\mu_y}}\\
&[\theta_{\mu_h},\sigma^2_{\mu_h}]\propto \{tr(\boldsymbol U_{\mu_h}^2)-\frac{1}{n_t}tr(\boldsymbol U_{\mu_h})^2\}^{1/2} \frac{1}{\sigma^2_{\mu_h}}
\end{aligned}
\end{equation*}
where $\boldsymbol U_{(.)}=\boldsymbol V_{(.)}^{-1}\frac{\partial \boldsymbol V_{(.)}}{\partial{\theta_{(.)}}}$ ; $n_t$ is the dimension of $\mu_y$(or $\mu_h$); $n_i$ is the dimension of $\psi_i$. The posteriors are proper when the common intercept estimate is avoided \citep*{berger2001objective}, whereas such propriety is not affected by individual intercepts.

For the individual Gaussian process $\boldsymbol \psi_i$'s, we use a combination of the Jeffreys prior and the hierarchical half-Cauchy prior:

\begin{equation*}
\begin{aligned}
&[\theta_{\psi}]\propto \{\sum_i tr(\boldsymbol U_{\psi_i}^2)\}^{1/2}\\ 
&{\sigma_{\psi_i}}\stackrel{indep.}{\sim} C^+ (0,\tau)
& \tau {\sim} C^+(0,\sigma_y)
\end{aligned}
\end{equation*}
For the nuisance parameter, we assume $[\sigma^2_y]\propto 1/\sigma^2_y$. It might seem tempting to also use Jeffreys prior on ${\sigma^2_{\psi_i}}$; however, this would lead to either under- or over-estimation of $\sigma^2_y$. If used, it would implicitly assume complete independence between $\sigma^2_{\psi_i}$'s and $\sigma^2_y$, whereas the two should be correlated in scale, as $\sigma^2_{\psi_i}$'s and $\sigma^2_y$ represent the last pieces of signals and the noise, respectively. The hierarchical parameter $\tau$ is necessary to prevent undesirable rigidity from the scaling of $\sigma_y$. This prior is also known as horseshoe prior and was proposed by \cite{carvalho2010horseshoe}.

For the intercept and intercorrelation coefficients, it seems ideal to assign flat prior $[.]\propto 1$. However, this results in unidentifiability of the model. To solve this issue, we introduce shrinkage with g-priors \citep*{zellner1986assessing}:

\begin{equation*}
\begin{aligned}
&\gamma_i \stackrel{indep.}{\sim} N(0, g_\gamma \sigma^2_y/n_i)\\
&\eta_i \stackrel{indep.}{\sim} N(0, g_\eta /n_i)\\
&\phi \sim N(\mu_\phi, g_\phi /(\sum_i\boldsymbol \psi_i'\boldsymbol \psi_i))
\end{aligned}
\end{equation*}
To make the g-prior as weakly informative as possible, we assign the Jeffreys prior to the hyperparameter  $[g_{(.)}]\propto 1/g_{(.)}$ and $[\mu_\phi]\propto1$. The scale parameters for $\eta_i$ and $\psi$ are omitted, due to the notion that logit link implicitly assumes a logistic distribution with the scale fixed at $1$.

\section{Simulation Studies}

\subsection{Estimation of the Latent Processes}

To demonstrate the accuracy of latent process estimation, we carried out the following simulation:

\begin{equation*}
\begin{aligned}
&\textbf{Latent processes:}\\
&\mu_y(x)=50 \sin(\frac{x-20}{100})\cos(-\frac{x-10}{15})\\
&\mu_h(x)=4 \sin(\frac{x-10}{5})\cos(\frac{x}{10})\\
&{\boldsymbol \psi}_i \stackrel{indep}{\sim} {N}({\boldsymbol 0},\boldsymbol V_\psi \sigma^2_{\psi_i}) \text{, $\boldsymbol V_\psi$ is of AR(1) and $\sigma_{\psi_i}\sim U(0.5,1)$}\\
&{\boldsymbol H}_i=\boldsymbol \mu_h+\phi\boldsymbol \psi_i\\
&{\boldsymbol \lambda_i}=\frac{exp({\boldsymbol H}_i)}{1+exp({\boldsymbol H}_i)}\\
&\textbf{Observed processes:}\\
&{\boldsymbol Y}_i \sim { N}( \boldsymbol  \mu_y+\gamma_i{\boldsymbol 1}+{\boldsymbol \psi}_i,  0.01{ \boldsymbol I})\\
&{\boldsymbol R}_i \sim { Bin} (\boldsymbol \lambda_i)
\end{aligned}
\label{sim1}
\end{equation*}

\begin{figure}[!H]

\begin{subfigure}[b]{.3\columnwidth}
\centering\includegraphics[width=1\columnwidth]{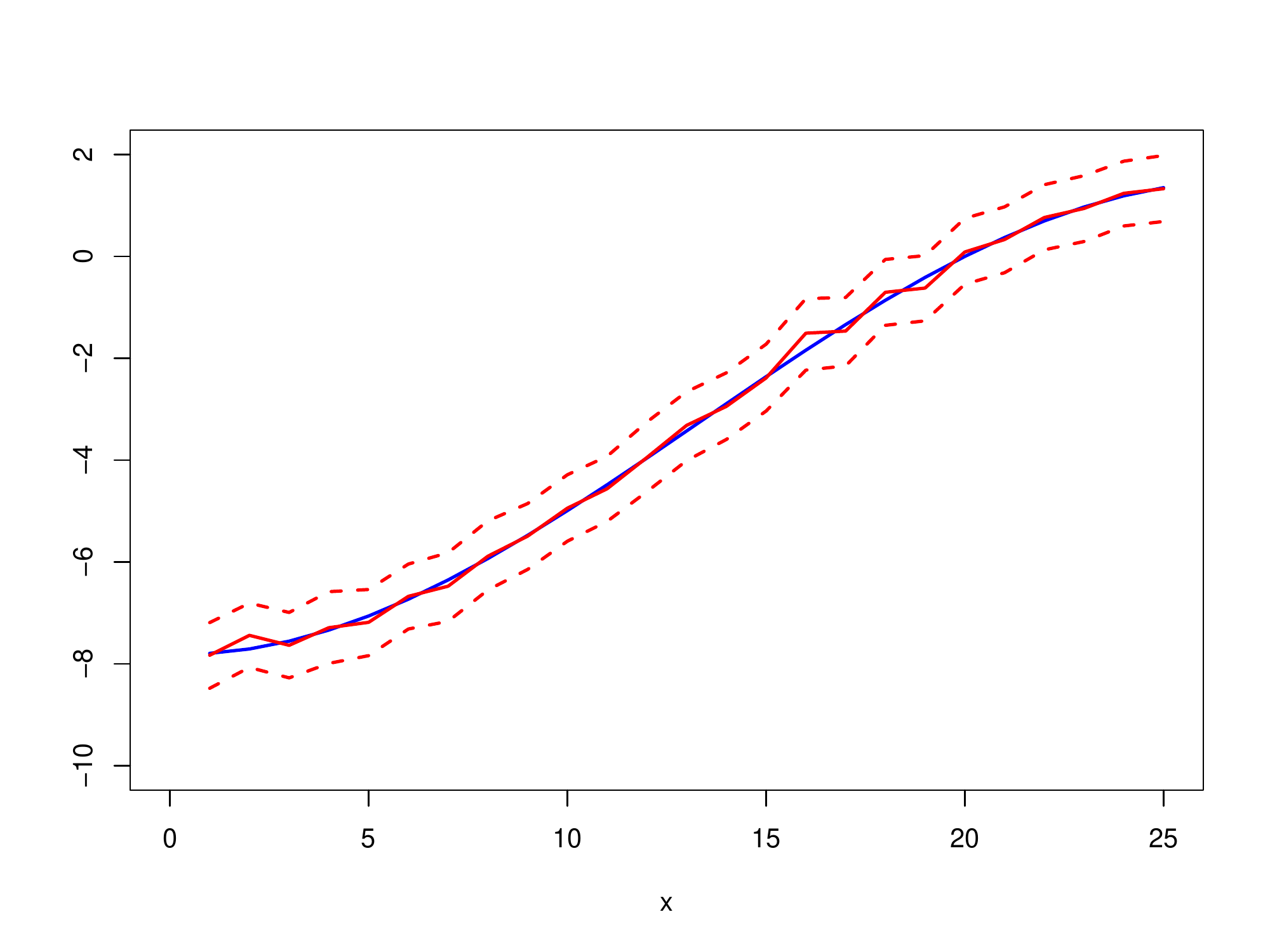}
\subcaption{$\mu_y$ in Sim No. 1}
\end{subfigure}
\begin{subfigure}[b]{.3\columnwidth}
\centering\includegraphics[width=1\columnwidth]{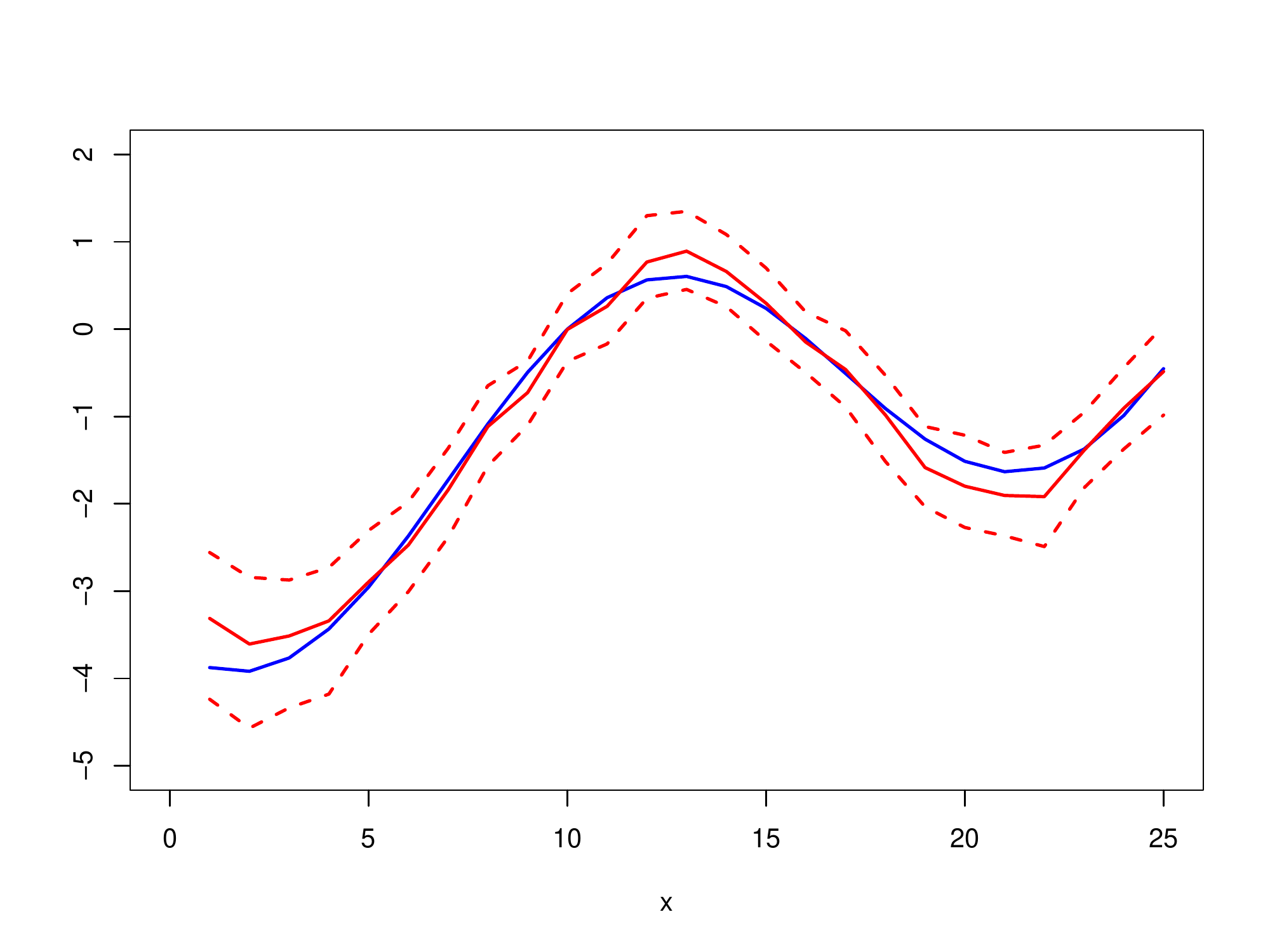}
\subcaption{$\mu_h$  in Sim No. 1}
\end{subfigure}
\begin{subfigure}[b]{.3\columnwidth}
\centering	\includegraphics[width=1\columnwidth]{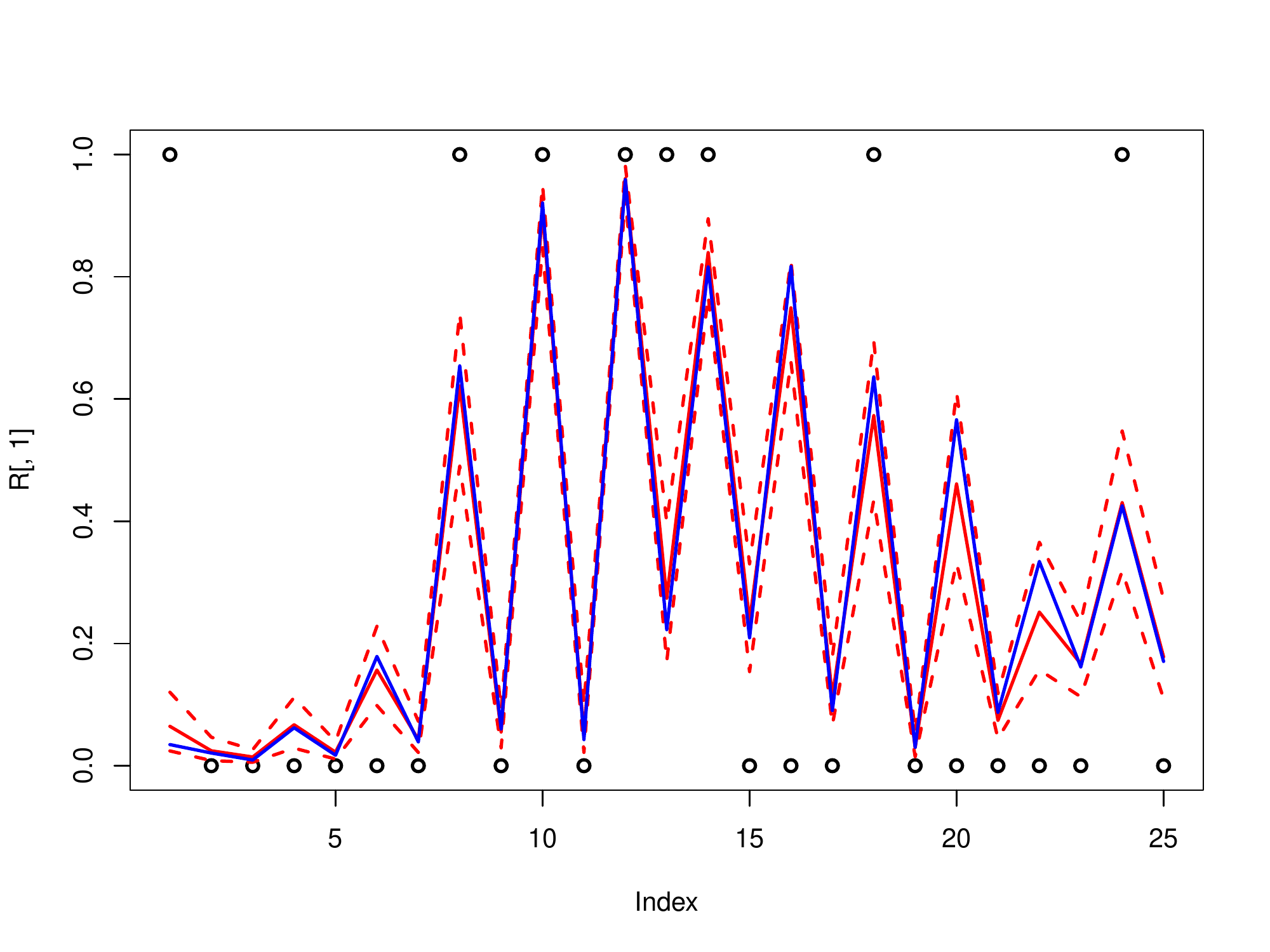}
\subcaption{$\lambda$  in Sim No. 1}
\end{subfigure}

\begin{subfigure}[b]{.3\columnwidth}
\centering\includegraphics[width=1\columnwidth]{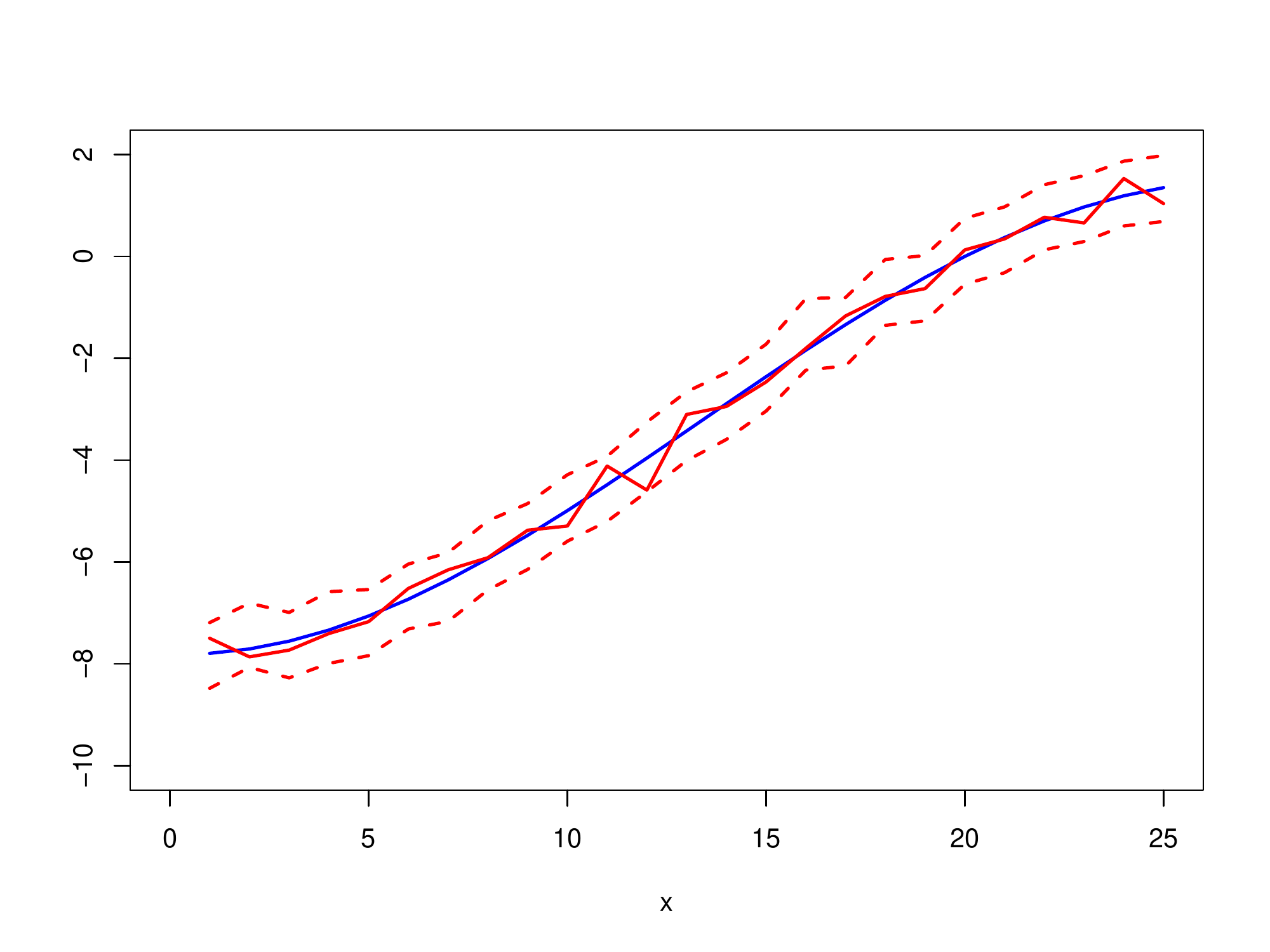}
\subcaption{$\mu_y$  in Sim No. 2}
\end{subfigure}
\begin{subfigure}[b]{.3\columnwidth}
\centering\includegraphics[width=1\columnwidth]{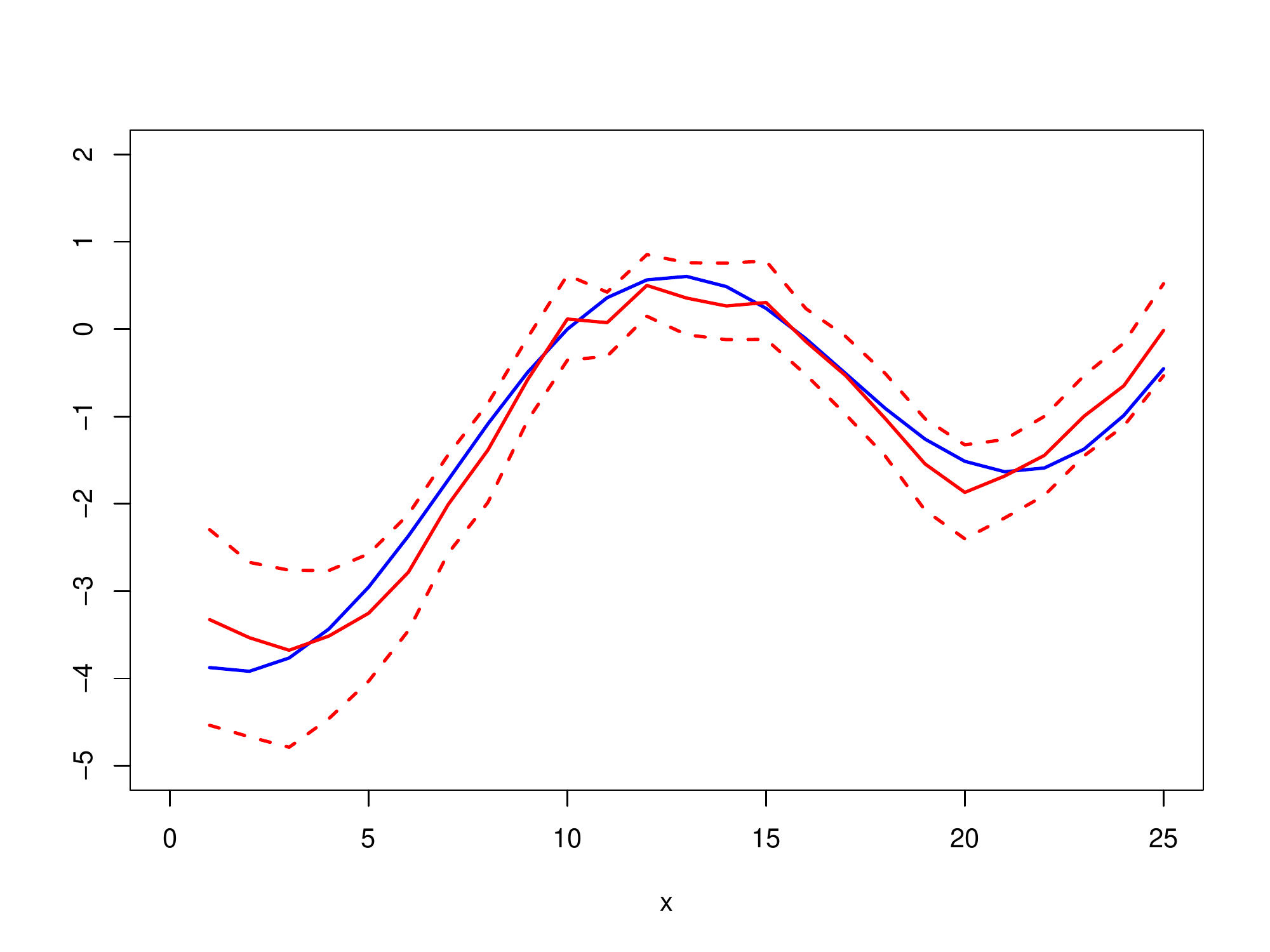}
\subcaption{$\mu_h$  in Sim No. 2}
\end{subfigure}
\begin{subfigure}[b]{.3\columnwidth}
\centering	\includegraphics[width=1\columnwidth]{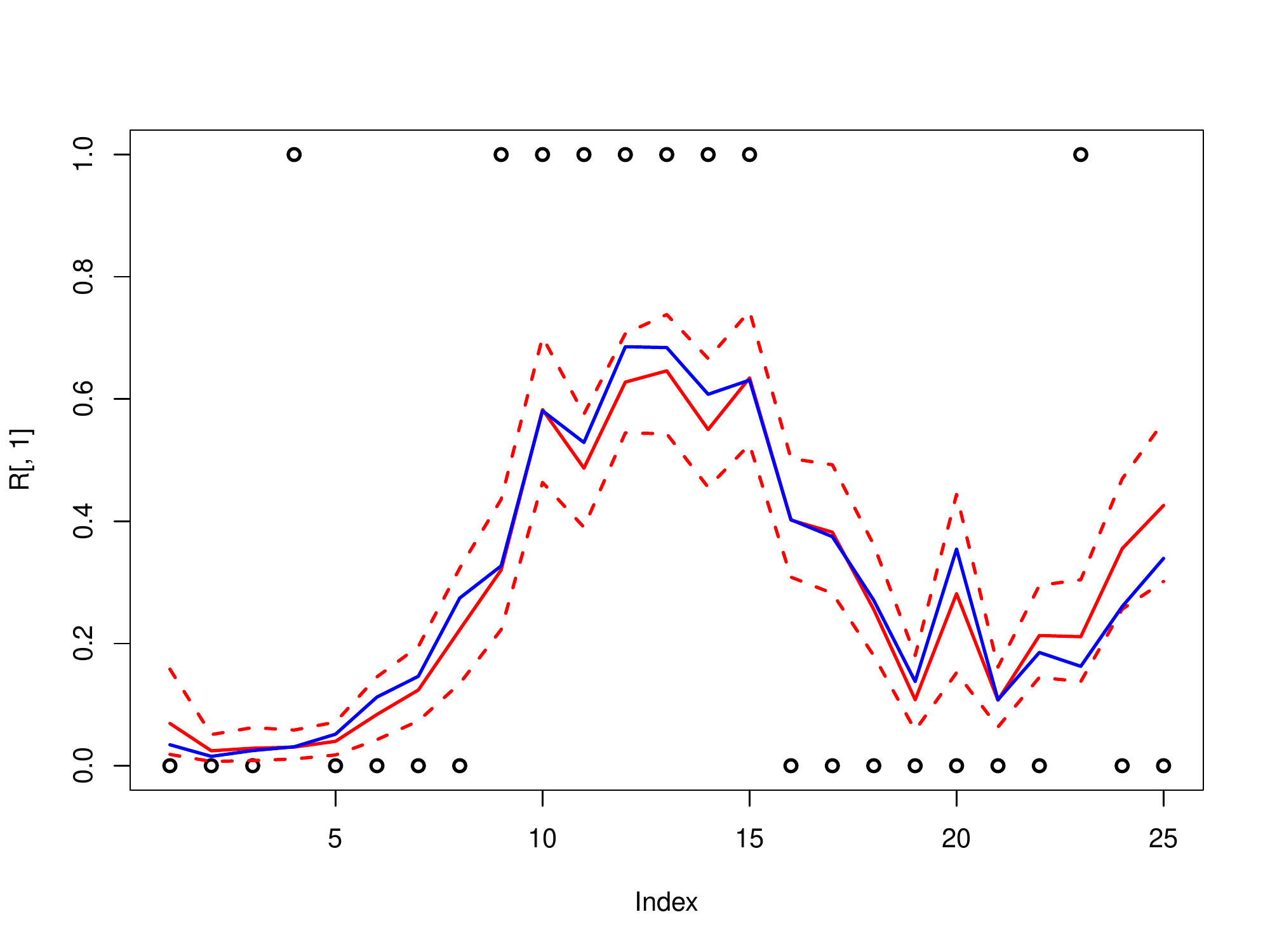}
\subcaption{$\lambda$  in Sim No. 2}
\end{subfigure}
			
\begin{subfigure}[b]{.3\columnwidth}
\centering\includegraphics[width=1\columnwidth]{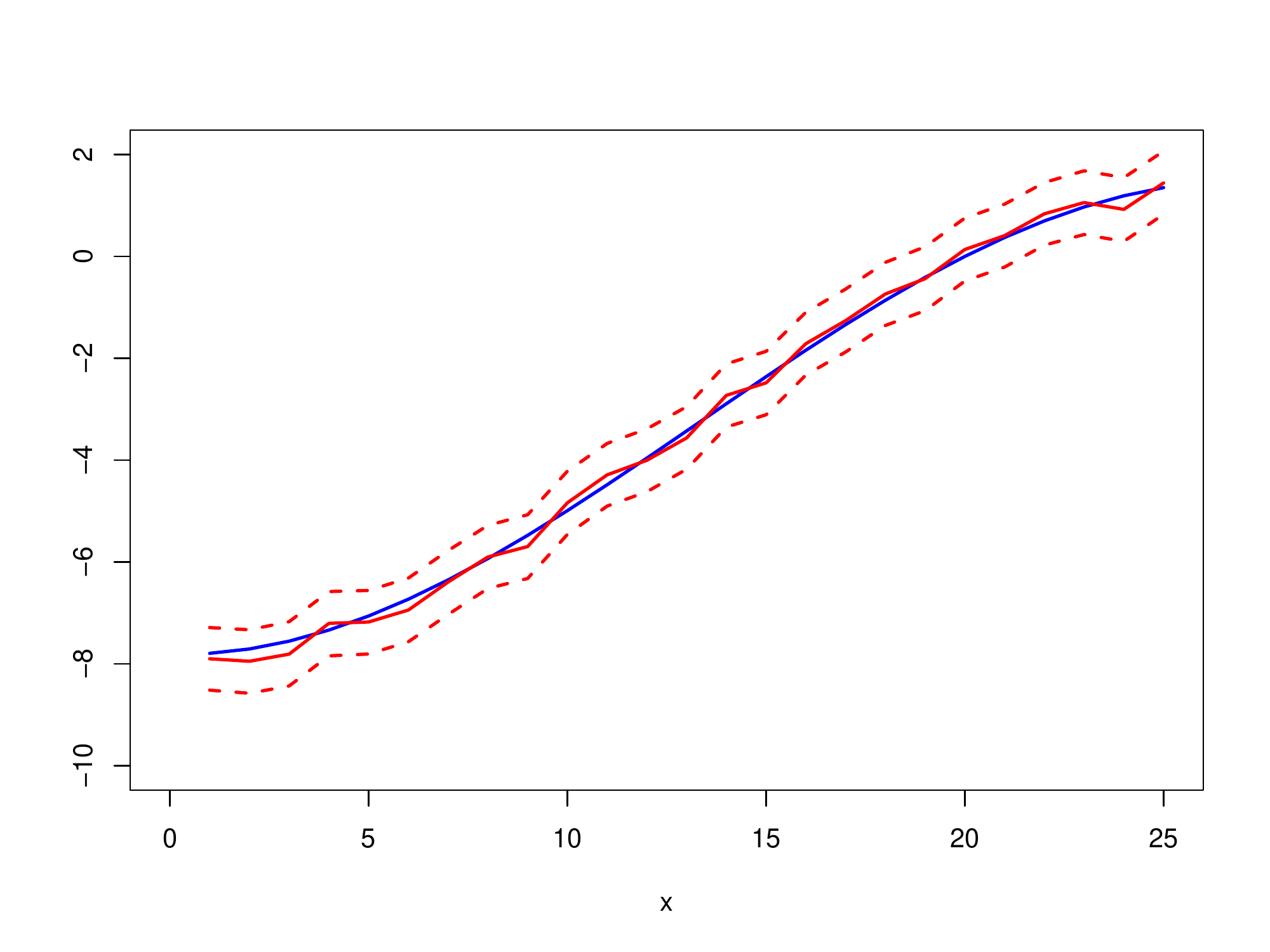}
\subcaption{$\mu_y$  in Sim No. 3}
\end{subfigure}
\begin{subfigure}[b]{.3\columnwidth}
\centering\includegraphics[width=1\columnwidth]{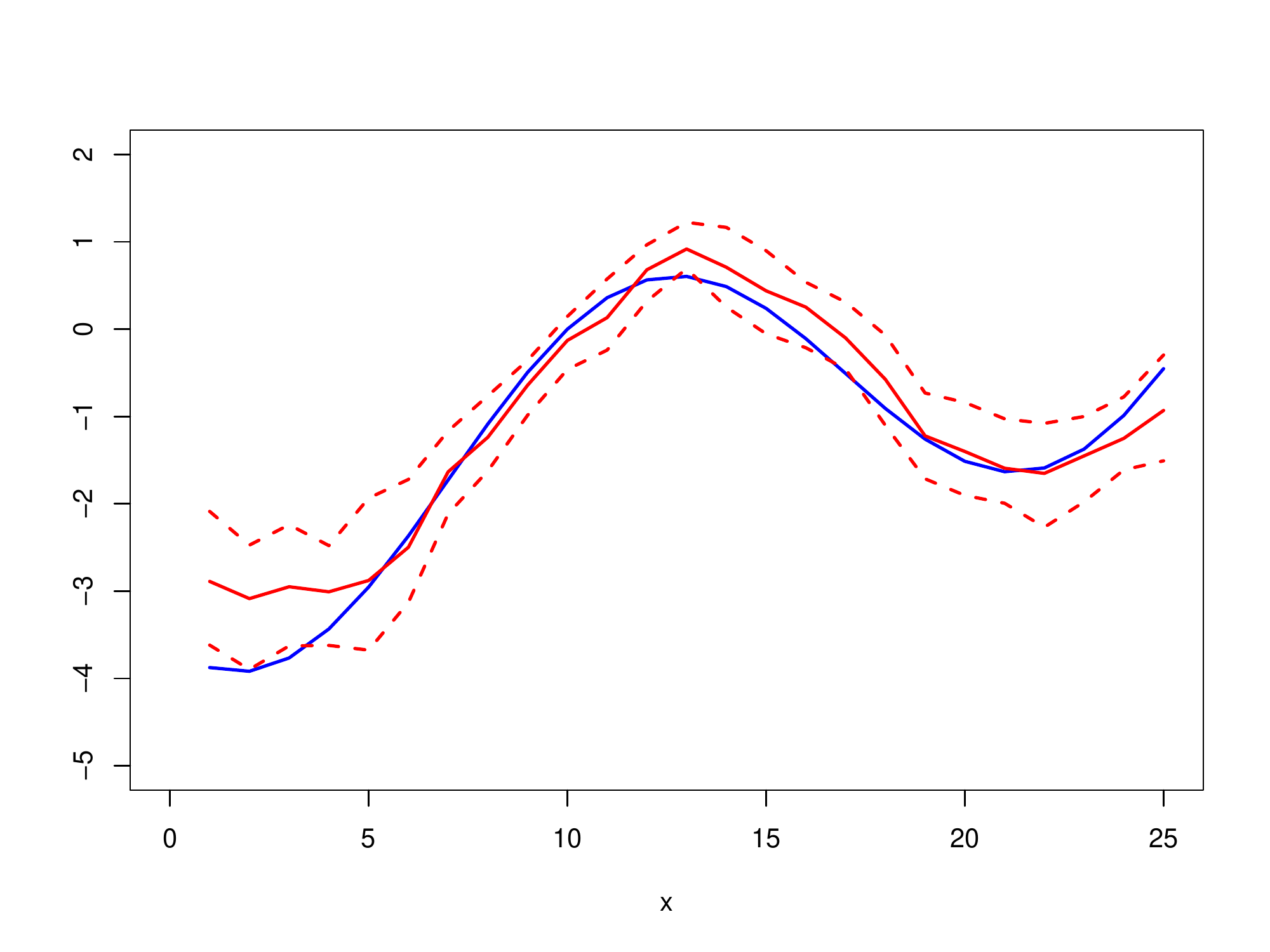}
\subcaption{$\mu_h$  in Sim No. 3}
\end{subfigure}
\begin{subfigure}[b]{.3\columnwidth}
\centering	\includegraphics[width=1\columnwidth]{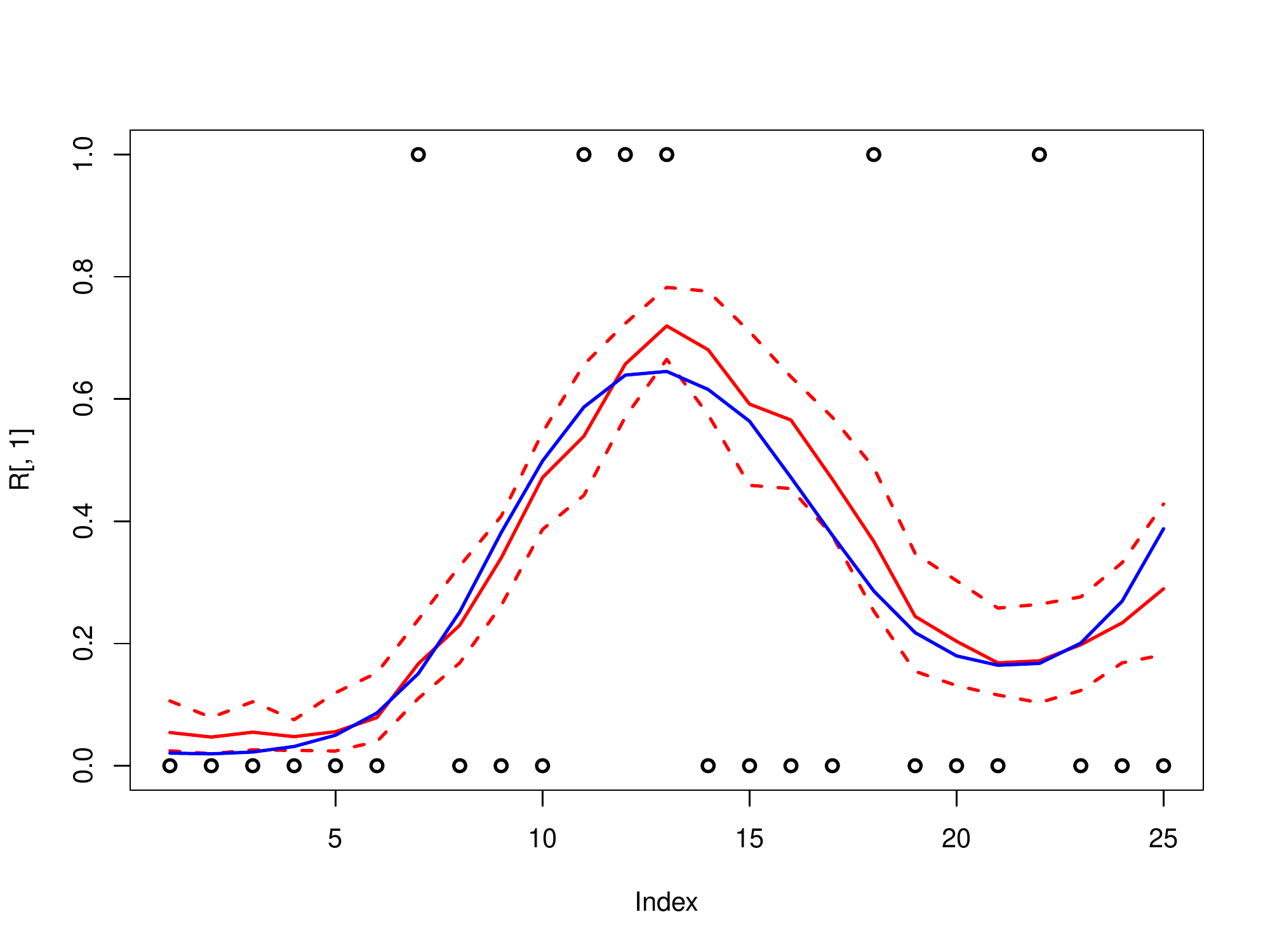}
\subcaption{$\lambda$  in Sim No. 3}
\end{subfigure}
			
\caption{Estimation of the latent processes using JHGP model in simulation studies. The true unknown processes are shown in blue, and the estimated values and the 95\% pointwise credible intervals are shown in red.}
\label{esti_sim}
\end{figure}

We generated the samples with three different sets values of $(\theta_\psi,\phi)$, which corresponds to different levels of association. To demonstrate the robustness of model to small sample sizes, we did the following for each setting. We simulated only 50 subjects ($i=1,...,50$), each with 25 time points ($x=1,...,25$). We fit the JHGP model to the three sets of data. The estimation of parameters are shown in Table~\ref{sim_par_est}; plots of latent process are shown in Figure~\ref{esti_sim}. The model correctly identified the values of autocorrelation $\theta_\psi$ and the association parameter $\phi$. Moreover, the nonlinear latent hierarchies $\mu_y$ and $\mu_h$ were both accurately estimated. The hazard function estimates for $\lambda$, which are hidden behind the binary outcomes ${\boldsymbol R}$, show high correlation with the true values. Therefore, we conclude that our JHGP model is robust to different parametrizations.

\begin{table}[ht!]
 \begin{center}
\begin{tabular}{ l  | c c}
\hline                        
Sim No. (true values)&				$\theta_\psi$ & $\phi$ \\ 
\hline                        
Sim 1 ($\theta_\psi=-0.8$, $\phi=0.9$)   &  -0.77 (-0.81,-0.74) &   0.86 (0.69, 1.04)  \\ 
Sim 2 ($\theta_\psi=-0.5$, $\phi=-0.3$ )& -0.53 (-0.48,-0.57) &   -0.28 (-0.44, -0.12)  \\
Sim 3 ($\theta_\psi=-0.1$, $\phi=0.01$ ) &  -0.09 (-0.14,-0.02) &   0.03 (-0.10, 0.18)   \\
\hline                        
\end{tabular}
\end{center}
\caption{Estimation of parameters with different $(\theta_\psi,\phi)$. The posterior means (with 95\% credible intervals) are shown. }
\label{sim_par_est}
\end{table}

\subsection{Choice of Covariance Function for Individual Process}

The choice of covariance function affects the behavior of extrapolation curve. In the population hierarchy estimates ($\boldsymbol \mu$), the results do not seem to differ much by covariance selection (except for differentiabilities). On the individual level (${\boldsymbol Y}_i$ and ${\boldsymbol H}_i$), the basic properties of the chosen covariance function are directly exhibited its prediction mean.

\begin{figure}[!ht]
 
          \begin{subfigure}[b]{.4\textwidth}
          \begin{center}
\includegraphics[width=\linewidth]{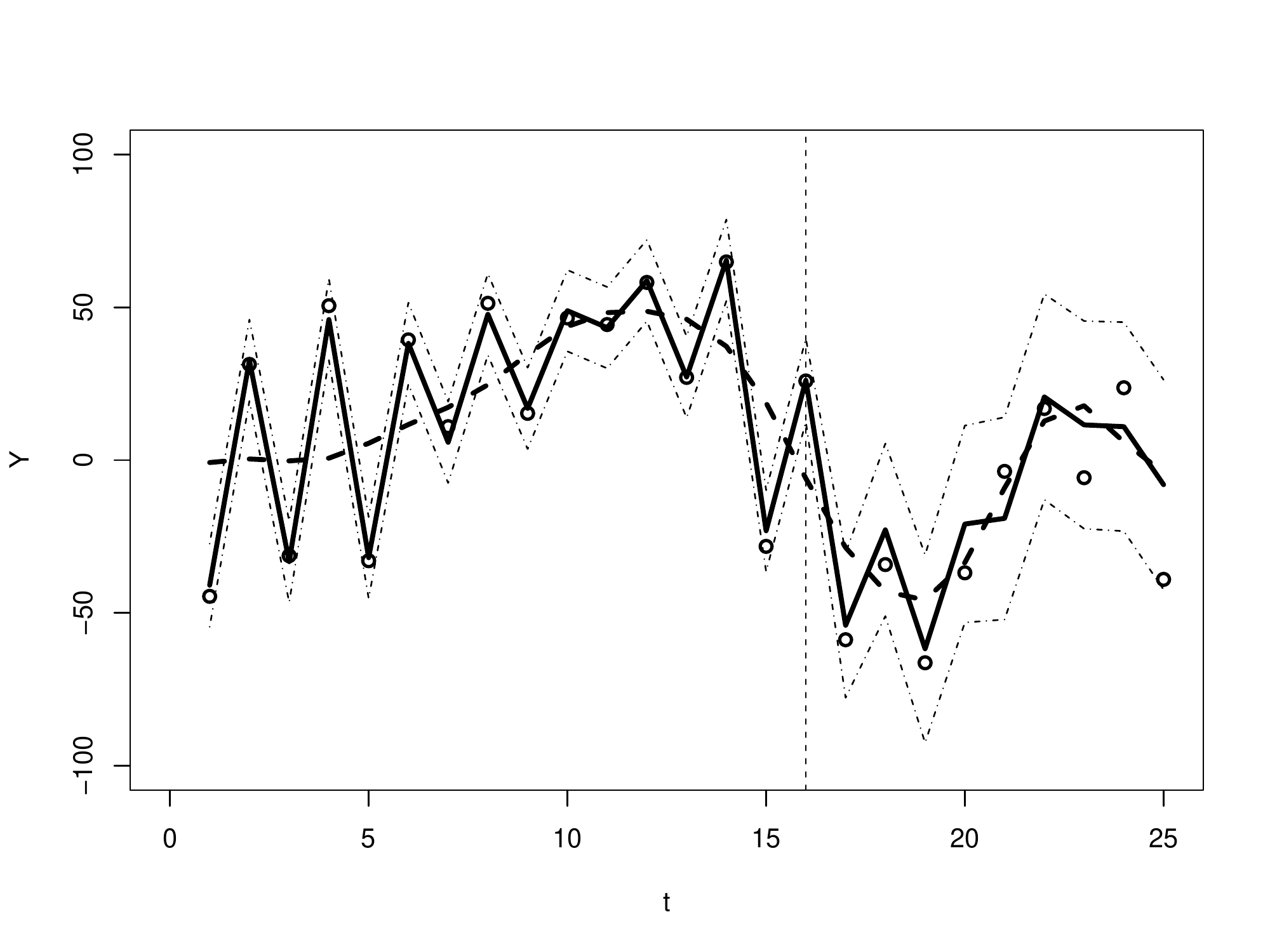}
		\end{center}
				\caption{AR(1) GP.}
		\end{subfigure}
		\begin{subfigure}[b]{.4\textwidth}
		          \begin{center}
\includegraphics[width=\linewidth]{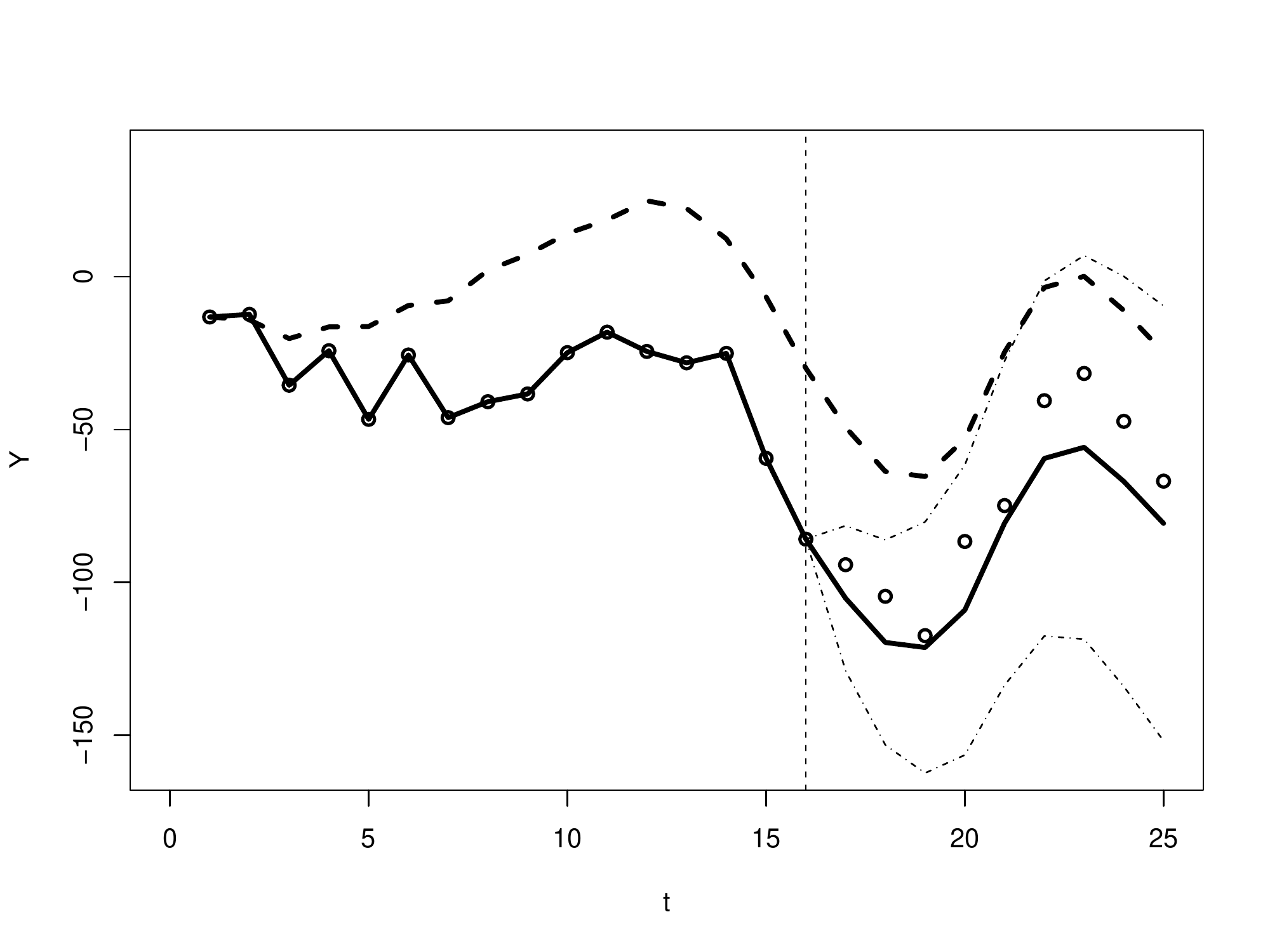}
		\end{center}
				\caption{Brownian motion GP}
		\end{subfigure}
\caption{Different types of covariance functions in the individual process leads to different extrapolation curves (solid lines on the right).}
\label{cov_compare}
\end{figure}

We conducted a simple comparison between a stationary and non-stationary covariance function . As shown in Figure \ref{cov_compare}, the stationary AR(1) process with a negative $\theta_\psi$  tends to oscillate around the mean. This property is useful if the subject trajectory is expected to progress similarly to other. On the other hand, Brownian motion as a martingale process always shows a constant difference from the mean process. This can reflect the notion that a loss or gain at a certain time is permanent for an individual. Such properties can be used together by choosing the sum of two different covariances.

\subsection{Signal Detection of Association}

In the JHGP model, the association between two responses is established by the shared individual Gaussian process. Its strength can be measured by $\phi$ in Eqn~\ref{jhgp_cond}. We assessed the effectiveness of $\phi$ in association detection under various levels of interference.

We use the same equations in Sec 3.1 to generate test samples with $\mathbb{\phi}=0.5$, except we add a noise vector $ {\boldsymbol \tau}_i\sim  N( \boldsymbol 0,  \boldsymbol I\sigma^2_\tau)$ to ${\boldsymbol H}_i$.

\begin{align*}
&{\boldsymbol H}_i=\boldsymbol \mu_h+\phi\boldsymbol \psi_i+ {\boldsymbol \tau}_i
\end{align*}

 We then gradually increase $\sigma^2_\tau$ in order to disturb the estimation of $\mathbb{\phi}$. The noise-signal ratio is controlled by $\sigma^2_\tau / ( \phi^2  ||\sigma^2_{\psi_i}||)$, where $ ||\sigma^2_{\psi_i}||$ is the average of $\sigma^2_{\psi_i}$. The results are shown in Table~\ref{signal_detect}. The JHGP model exhibits robustness in the presence of disturbance. The numerical estimates only start to degrade around noise-signal ratio of $8.0$ yet the association remains significant until the magnitude reaches $32.0$. We conclude that the JHGP model is very robust in detecting the association between two responses.

\begin{table}[ht!]
\caption{Association measures under different noise levels} \label{signal_detect}
 \begin{center}
\begin{tabular}{ c  | c  }
\hline                        
Noise/Signal ($\sigma^2_\tau / \phi^2 \sigma^2_\psi$) &$\phi$ (true value: 0.50)  \\
\hline 
$0.1$   &  0.50 (0.34, 0.65)    \\                       
$0.5$   &  0.46 (0.29, 0.58)   \\                   
$1.0$   &  0.51 (0.37, 0.66)  \\                   
$2.0$   &  0.44 (0.30, 0.58)  \\                   
$4.0$   &  0.45 (0.30, 0.60)    \\
$8.0$ & 0.25 (0.13, 0.37)	\\
$16.0$ & 0.23 (0.11, 0.36)   \\
$32.0$ & 0.09 (-0.03, 0.22)   \\
\hline                        
\end{tabular}
\end{center}
\end{table}

\subsection{Sensitivity-Specificity Studies}

\begin{figure}[!ht]
\begin{center}
\includegraphics[width=0.3\columnwidth]{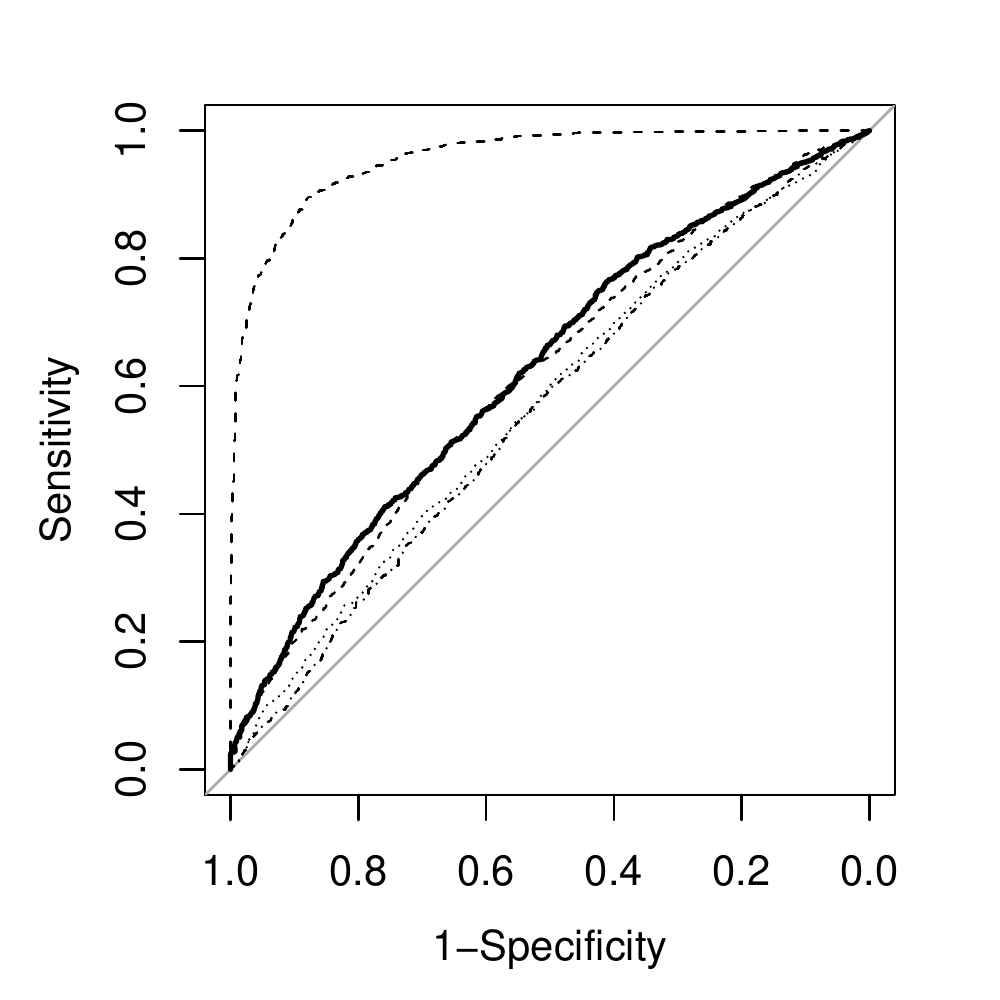}
\end{center}
\caption{Sensitivity-specificity analyses in simulation studies. From up-left to the diagonal, ROC curves (AUC) of model fitted with: JHGP (0.828), extended HGP(0.782), logistic regression (0.626).}
\label{ROC_sim}
\end{figure}

We conduct sensitivity analysis on the survival part of the JHGP model. We compare the results using JHGP, HGP and simple logistic regression. The posterior means of $\lambda_i$ are used as the fitted probabilities in the first two models. As shown in Figure \ref{ROC_sim}, the JHGP model has largest area under curve measure ($AUC=82.8\%$), while the HGP model is weaker  ($AUC=78.2\%$). This supports the notion that joint modeling provides better estimation for the hazards. The least favorable model is the simple logistic regression ($AUC=62.6\%$), in which ${Y}_i$ is treated as a covariate.

\subsection{Forecasting Performance}

We censor each subject in the simulated data using random time $C_i=\min(\max({X}_i),t_c)$ where $t_c\sim U(0, 2\max({X}_i))$ and $\max({X}_i)$ is the last recorded time in that subject. This mechanism results in censoring in about $50\%$ of the subjects, for which censoring occurs at random time points.

 \begin{figure}[!H]
          \begin{subfigure}[b]{.4\columnwidth}
          \begin{center}
\includegraphics[width=\linewidth]{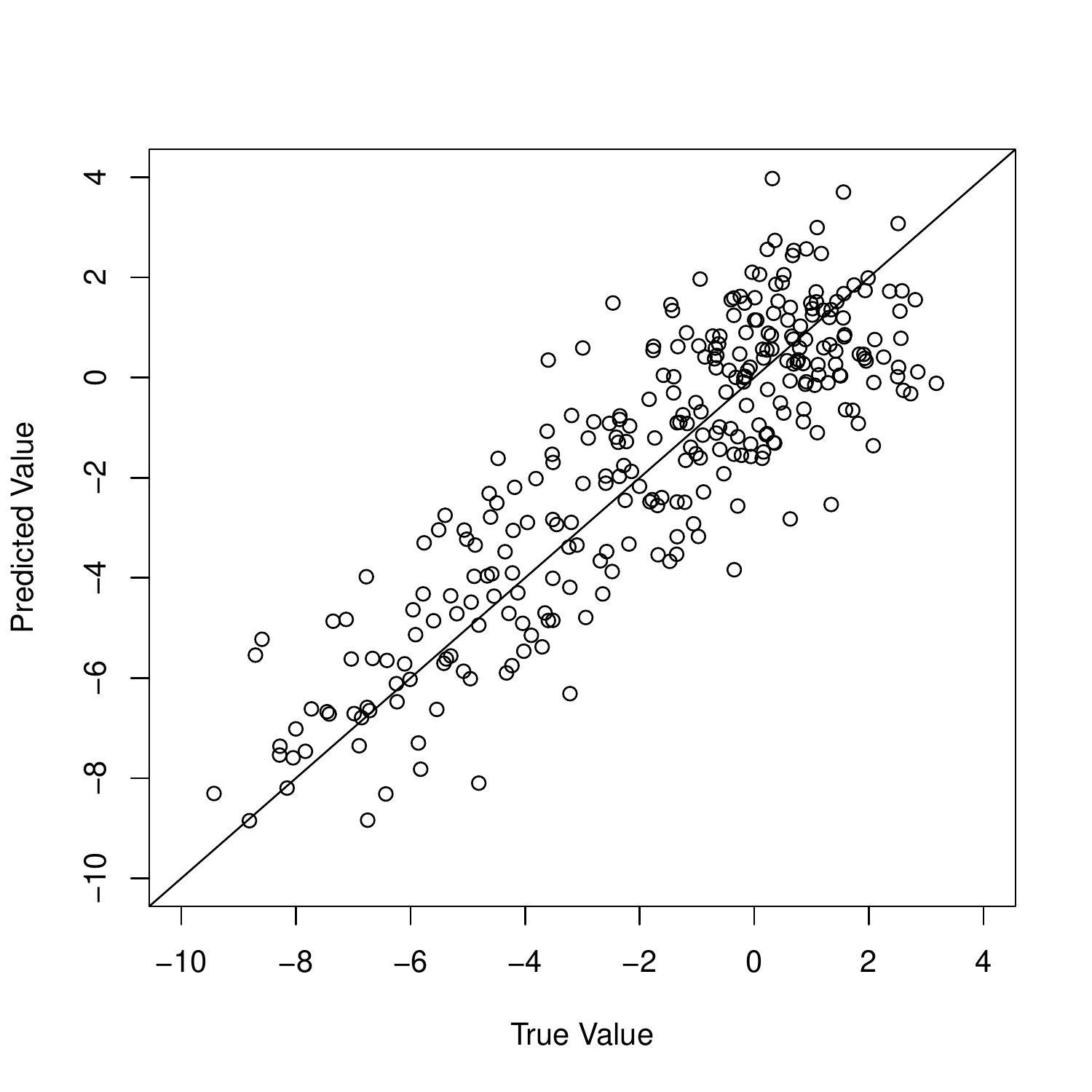}
		\end{center}
		\caption{Predicted vs true  $ Y$}
		\end{subfigure}
          \begin{subfigure}[b]{.4\columnwidth}
          \begin{center}
\includegraphics[width=\linewidth]{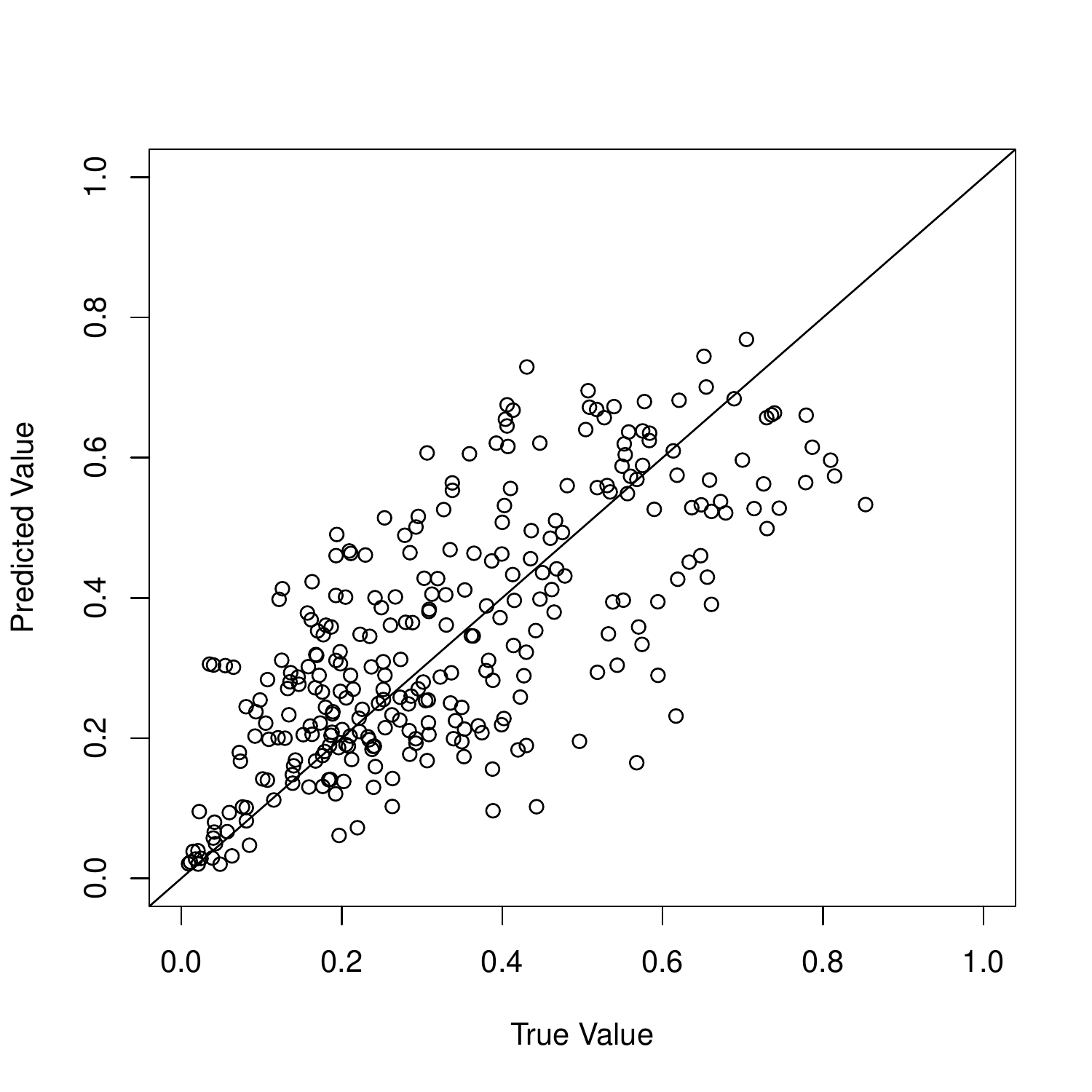}
		\end{center}
		\caption{Predicted vs true $ \lambda$}
		\end{subfigure}
\caption{Forecasting performance in simulation studies. Comparison plots of the predicted vs the true values.}
\label{pred_sim}
\end{figure}

\begin{table}[H]

\begin{tabular}{ l  c  c}
\hline
&$ Y$ &$ \lambda$\\
\hline                  
MPSD & 0.25 & 0.19  \\
MAD & 0.64 &  0.30 \\
RMSE & 0.52 & 0.71 \\
Cor & 0.86 & 0.73\\
\hline  
\end{tabular}
\caption{Forecasting performance of JHGP model in simulation studies. Mean posterior standard deviation (MPSD), median absolute deviation (MAD), Root Mean Square Error (RMSE) and Pearson correlation (Cor) are shown. The first three metrics are shown in relative magnitudes, as compared with absolute posterior mean, median absolute value and standard deviation.}
\label{pred_sim_table}
 \end{table}

The assessment of the forecasting performance is presented in Table~\ref{pred_sim_table} and illustrated in Figure~\ref{pred_sim}. The model shows high prediction precision, low bias and small prediction error. The predicted values are highly correlated to the true values.

\section{Application in Medical Monitoring}

The JHGP model is now applied to the motivating clinical problem. Percentage of forced expiratory volume in 1 second (FEV$_1\%$) is a common measure of lung function in cystic fibrosis (CF) patients. Studies have demonstrated that the rates of change differ in adolescence and adulthood  \citep*{vandenbranden2012lung} and  the decline of is nonlinear\citep*{szczesniak2013semiparametric}. Pulmonary exacerbation (PEx) is a temporary worsening of lung condition due to infection or inflammation and can occur multiple times in an individual CF patient. Therefore PEx needs be modeled as a recurrent survival event.
A previous study has also established an association between PEx and subsequent FEV$_1\%$ decline  \citep*{sanders2011pulmonary}. Patient-specific maximum quarterly FEV$_1\%$ and occurrence of PEx are used for the analysis. Data were acquired from the Cystic Fibrosis Foundation Patient Registry. The quarterly ages are used as the time indices for the discrete model.  Among patients who have experienced both PEx and FEV$_1\%$ decline, we selected a sample of 38 subjects with 818 entries of observation. Then, the more recent 50\% of observations (both FEV$_1\%$ and PEx) are masked in 19 randomly chosen subjects. This subset results in a training and testing split of about 75\% and 25\%, respectively.

\begin{figure}[!H]
          \begin{subfigure}[b]{.4\textwidth}
          \begin{center}
\includegraphics[width=\linewidth]{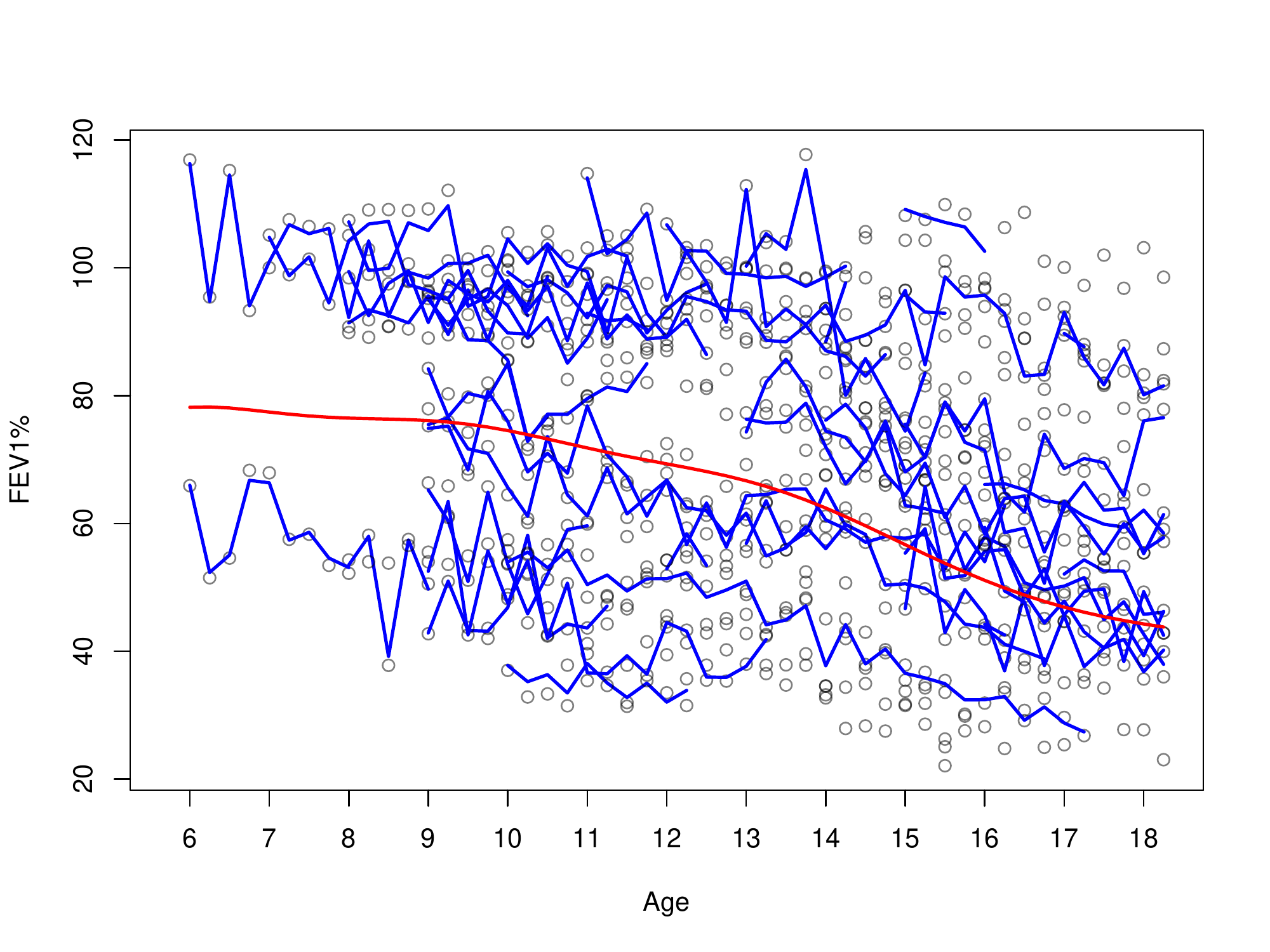}
		\end{center}
		\caption{FEV$_1\%$}
		\end{subfigure}
          \begin{subfigure}[b]{.4\textwidth}
          \begin{center}
\includegraphics[width=\linewidth]{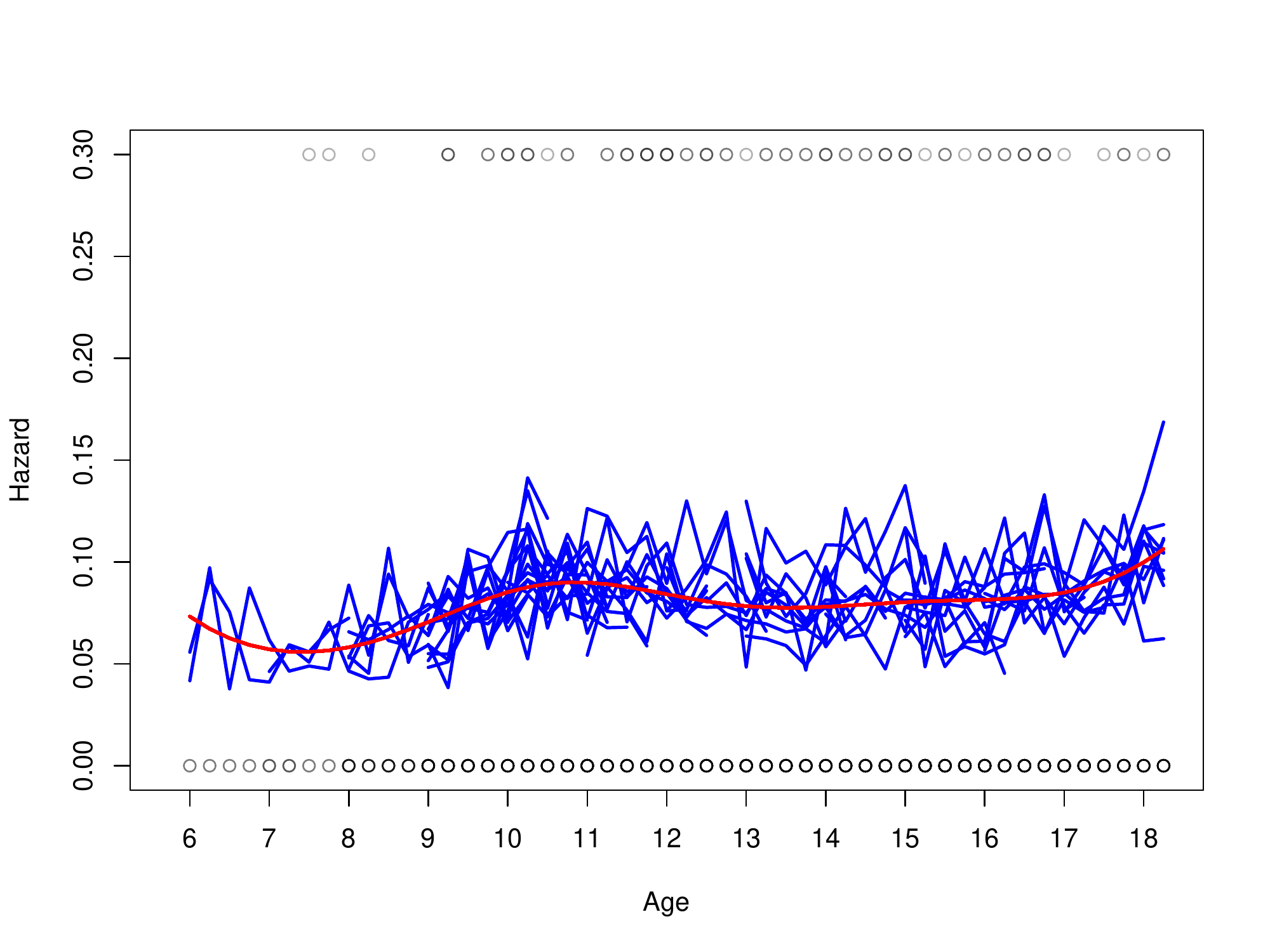}
		\end{center}
		\caption{PEx hazard}
		\end{subfigure}
	
\caption{Fitted values of FEV$_1\%$ and PEx hazard with JHGP model on CF data}
\label{data_fitted}
\end{figure}

We first focus on the parameter estimation.The JHGP model detects a strong autocorrelation ($\theta_\psi=-0.82$) in the shared Gaussian process $\psi$; the variations of FEV$_1\%$ and PEx hazard has a negative correlation ($\phi=-0.11$). It is worth mentioning that the estimate of $\phi$ indicates a strong association. The small magnitude is due to the fact that the FEV$_1\%$ has its mean around 70, while hazards are commonly limited to $(-10,10)$ under logit link. One possible way to increase the sensitivity of this parameter is to standardize $\boldsymbol Y$ before fitting the model; however, we kept FEV$_1\%$ on its original scale for the ease of clinical interpretation. The fitted FEV$_1\%$ and PEx hazard are shown in Figure~\ref{data_fitted}. The mean smoother and individual AR(1) processes are satisfactorily estimated. Among which, the population estimates are consistent with what we found in an earlier study using penalized splines \citep*{szczesniak2013semiparametric}. The baseline hazard has smooth estimates yet does not resemble any common parametric distribution, which indicates the flexibility in the Gaussian process. The stochasticity of individual variation in the PEx hazard is also captured by the JHGP model, due to the significant value of $\phi$. Caution is needed to assess the hazard function estimates at the two ends, where data are sparse and the estimation may be biased.

\begin{figure}[!H]

          \begin{center}
\includegraphics[width=0.3\linewidth]{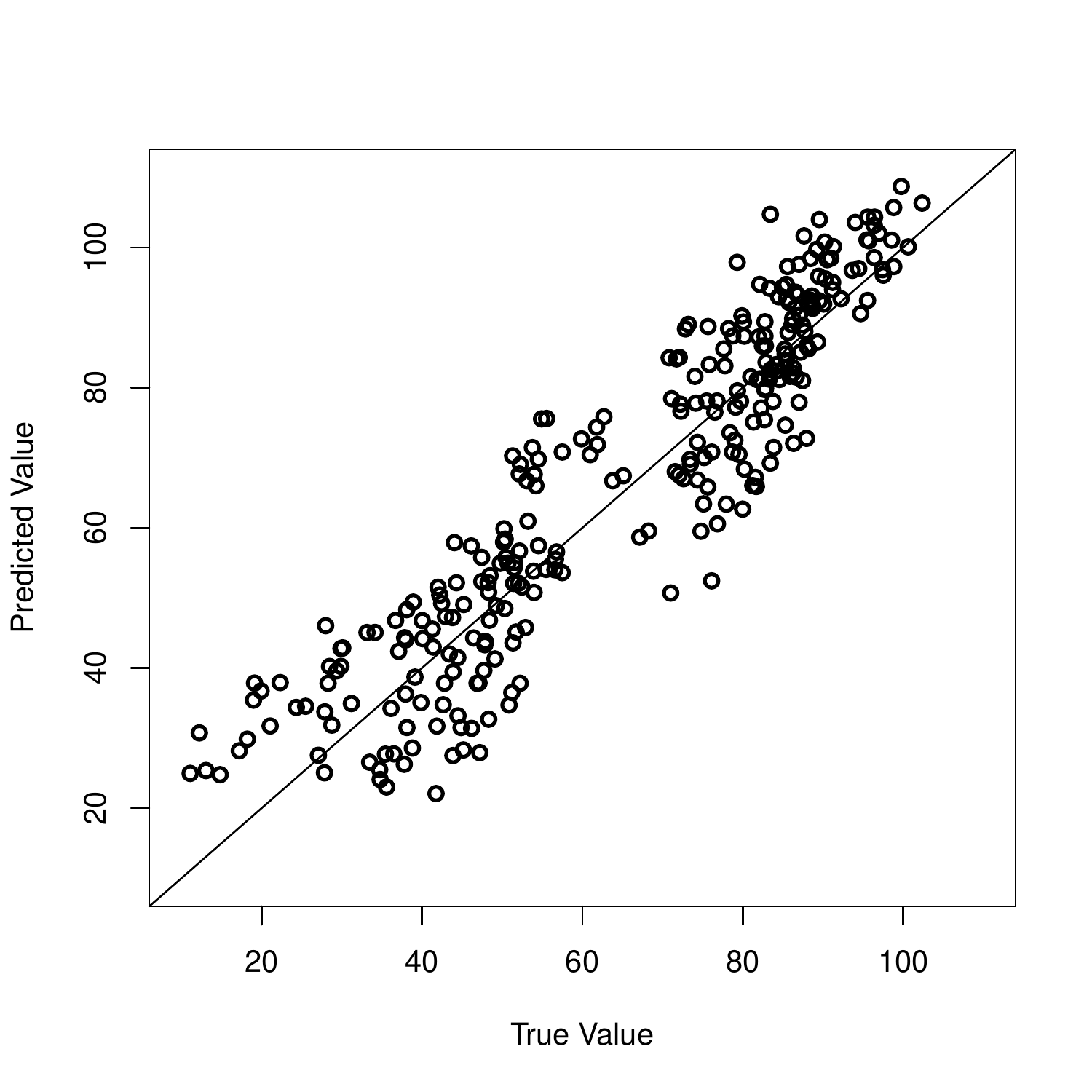}
		\end{center}

		\caption{Forecasting performance in FEV$_1\%$ of CF data. Comparison plots of the predicted vs the true values.}
		\label{data_true_pred}
		\end{figure}

\begin{figure}[!H]

\begin{center}
\includegraphics[width=.4\linewidth]{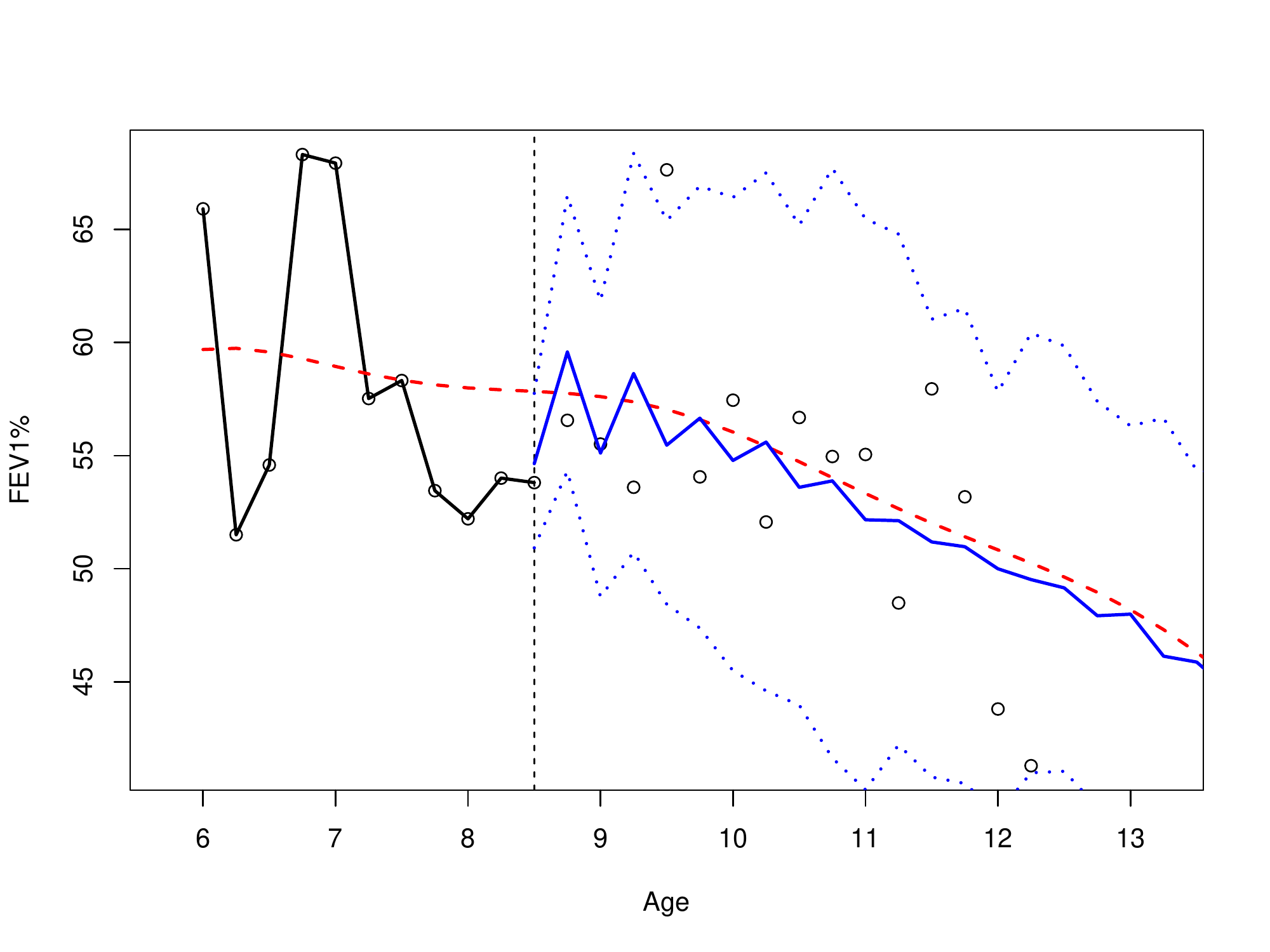}
\end{center}

\caption{Forecasting in FEV$_1\%$ using two hierarchies of Gaussian process. The population smoothed line (adjusted with individual intercept) are shown in red; and individualized AR(1) prediction is shown in blue. The 95\% credible intervals are also included.}
\label{data_predict_visual}
\end{figure}

We next analyze the forecasting performance of the JHGP model in FEV$_1\%$. The validation metrics are shown in Table~\ref{data_metrics}. Overall, the JHGP model shows high precision and low bias in forecasting (Figure~\ref{data_true_pred}). We further dissect the results and study the effects of the two hierarchies. The population Gaussian process seems adequate for predicting the future trend; however, the accuracy is further improved with the second individual process. Besides the better metrics, the improvement is illustrated in Figure~\ref{data_predict_visual}, where AR(1) process captures more details in the data.

\begin{table}[H]
 \begin{center}
\begin{tabular}{ l  c  c}
\hline                        
  & Population GP & JHGP \\
\hline                        
MPSD & -&  5.46  \\
MAD & 8.66 & 6.43   \\
RMSD & 10.63 & 8.47  \\
Cor & 0.89 & 0.93  \\
\hline  
\end{tabular}
\end{center}
 \caption{Forecasting performance in FEV$_1\%$ using JHGP model. To show the improvement of prediction from the individual hierarchy, the metrics of the population Gaussian process is also listed. The metrics are mean posterior standard deviation (MPSD), median absolute deviation (MAD), root mean square error (RMSE) and Pearson correlation (Cor) are shown. The metrics are shown in absolute magnitudes.} \label{data_metrics}
\end{table}

\begin{figure}[!ht]
\begin{center}
\includegraphics[width=.3\linewidth]{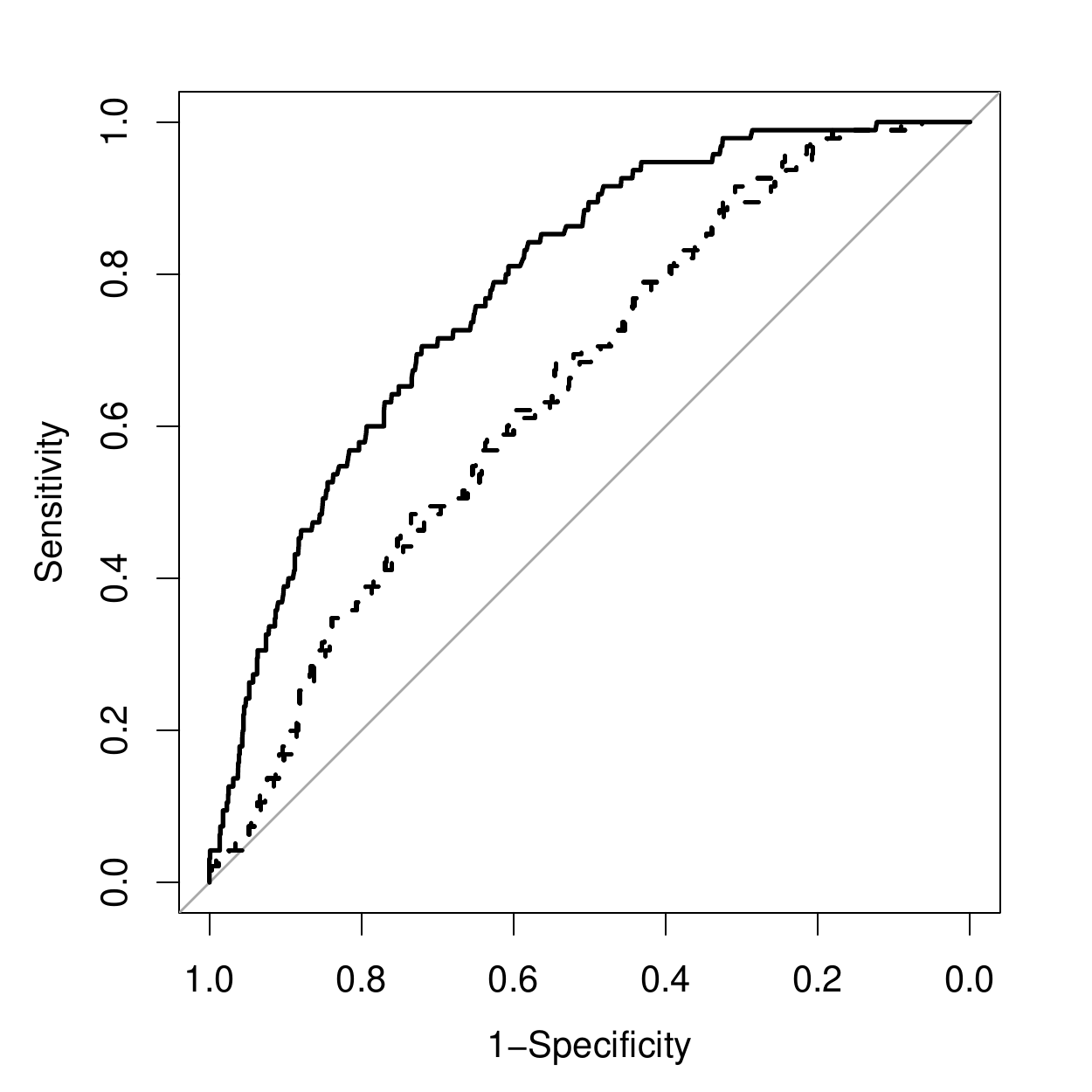}
\end{center}
\caption{ROC curves of fitting using CF data: JHGP (blue) shows higher AUC at 0.735, followed by extended HGP (red 0.683) and simple logistic regression (black 0.605).}
\label{ROC_data}
\end{figure}

Lastly, we study the sensitivity of the survival submodel in the JHGP. Similar to the simulation studies, we compare the ROC plot of the JHGP model with the extended HGP model and the simple logistic model with $FEV_1\%$ as a covariate (Figure~\ref{ROC_data}). The JHGP and extended HGP models show clear advantage over the traditional logistic model, probably due to their nonparametric nature. The consideration of joint modeling also boosts the sensitivity in comparison between the JHGP and extended HGP.

\section{Conclusion and Discussion}

We propose a novel hierarchical model that aims to accommodate the needs of subject forecasting. As a nonparametric approach, Gaussian process modeling has several advantages over traditional methods such as spline-based approaches. Most notably, the use of covariance function instead of knots enables automatic and robust estimation. This method has been widely used in machine learning  \citep*{rasmussen2006gaussian} and spatial statistics  \citep*{cressie1988spatial}. We further improve the Gaussian process approach with a two-hierarchy design: the first smooth Gaussian process describes the overall progression of data; the second stochastic Gaussian process captures the finer and personalized variation. The hierarchical design is not only conceptually clear, but also in alignment with one of the goals in longitudinal analysis: combining information from the levels of population and individuals.

As a predictive model, the JHGP shows high accuracy in the results of forecasting. We provide a flexible framework to incorporate the similarity in longitudinal data and the self-memory in time series analysis. The AR(1) structure can be easily replaced with more complex structures, as long as the covariance matrix can be derived. One possible issue may be the restriction on the positive definiteness of the covariance, however, since the population matrix is positive definite and has larger magnitude, this restriction may be lifted after adding the two variance matrices after reparameterization.

As a joint model, the JHGP adopts the individual fluctuations as the shared parameter. From the view of survival modeling, this parameter can be treated as the time-dependent frailty. We show that this design is robust to noise perturbation and also increases the sensitivity-specificity measure.

We have developed a fully Bayesian solution to the computation problem of the proposed model. Major progress has been made on Bayesian Gaussian process models in various areas, such as the development of priors \citep*{berger2001objective} \citep*{daniels1999nonconjugate} and dimension reduction \citep*{banerjee2013efficient}. On the other hand, there is less attention on the hierarchical use of Gaussian processes, especially using multiple Gaussian process simulatenously. One of the relevant works in this field is the use of finite mixtures of Gaussian processes \citep*{shi2005hierarchical}. Our hierarchical model differs in the sense that it is an additive model instead of a mixture model; therefore, we focus on controlling the scales of different components through prior conditioning.
The shrinkage effects of g-priors and the coupling of the smaller Gaussian process with noise enables correct estimation of the latent components.

Several extensions to this work can be made for the improvement of the prediction. If the longitudinal data exhibit several distinct types of progression, then the population hierarchy can be replaced with a mixture of Gaussian processes mentioned previously \citep*{shi2005hierarchical}. If the mean of the hazard function is clearly away from $0.5$, then other skewed link functions such as the generalized extreme value distribution \citep*{wang2010generalized} may be incorporated. Another direction to improve forecasting may reside in the use of nonstationary covariance functions \citep*{paciorek2004nonstationary}.

\section*{Acknowledgements}
Funding: Cystic Fibrosis Foundation Research and Development Program (grant number R457-CR11) provided partial support to R.D.S. and full support to L.L.D.
The authors are grateful to the Cystic Fibrosis Foundation Patient Registry Committee for their thoughtful comments and data dispensation.
\bibliographystyle{plainnat}
\bibliography{reference}

\begin{thebibliography}{21}
\providecommand{\natexlab}[1]{#1}
\providecommand{\url}[1]{\texttt{#1}}
\expandafter\ifx\csname urlstyle\endcsname\relax
  \providecommand{\doi}[1]{doi: #1}\else
  \providecommand{\doi}{doi: \begingroup \urlstyle{rm}\Url}\fi

\bibitem[Banerjee et~al.(2013)Banerjee, Dunson, and
  Tokdar]{banerjee2013efficient}
Anjishnu Banerjee, David~B Dunson, and Surya~T Tokdar.
\newblock Efficient gaussian process regression for large datasets.
\newblock \emph{Biometrika}, 100\penalty0 (1):\penalty0 75--89, 2013.

\bibitem[Berger et~al.(2001)Berger, De~Oliveira, and
  Sans{\'o}]{berger2001objective}
James~O Berger, Victor De~Oliveira, and Bruno Sans{\'o}.
\newblock Objective bayesian analysis of spatially correlated data.
\newblock \emph{Journal of the American Statistical Association}, 96\penalty0
  (456):\penalty0 1361--1374, 2001.

\bibitem[Carvalho et~al.(2010)Carvalho, Polson, and
  Scott]{carvalho2010horseshoe}
Carlos~M Carvalho, Nicholas~G Polson, and James~G Scott.
\newblock The horseshoe estimator for sparse signals.
\newblock \emph{Biometrika}, 97\penalty0 (2):\penalty0 465--480, 2010.

\bibitem[Cox(1972)]{cox1972regression}
David~R Cox.
\newblock Regression models and life-tables.
\newblock \emph{Journal of the Royal Statistical Society. Series B
  (Methodological)}, 34\penalty0 (2):\penalty0 187--220, 1972.

\bibitem[Cressie(1988)]{cressie1988spatial}
Noel Cressie.
\newblock Spatial prediction and ordinary kriging.
\newblock \emph{Mathematical Geology}, 20\penalty0 (4):\penalty0 405--421,
  1988.

\bibitem[Daniels and Kass(1999)]{daniels1999nonconjugate}
Michael~J Daniels and Robert~E Kass.
\newblock Nonconjugate bayesian estimation of covariance matrices and its use
  in hierarchical models.
\newblock \emph{Journal of the American Statistical Association}, 94\penalty0
  (448):\penalty0 1254--1263, 1999.

\bibitem[DiMatteo et~al.(2001)DiMatteo, Genovese, and
  Kass]{dimatteo2001bayesian}
Ilaria DiMatteo, Christopher~R Genovese, and Robert~E Kass.
\newblock Bayesian curve-fitting with free-knot splines.
\newblock \emph{Biometrika}, 88\penalty0 (4):\penalty0 1055--1071, 2001.

\bibitem[Eilers and Marx(1996)]{eilers1996flexible}
Paul~HC Eilers and Brian~D Marx.
\newblock Flexible smoothing with b-splines and penalties.
\newblock \emph{Statistical science}, 11:\penalty0 89--102, 1996.

\bibitem[Paciorek and Schervish(2004)]{paciorek2004nonstationary}
Christopher~J Paciorek and Mark~J Schervish.
\newblock Nonstationary covariance functions for gaussian process regression.
\newblock \emph{Advances in neural information processing systems},
  16:\penalty0 273--280, 2004.

\bibitem[Polson et~al.(2012)Polson, Scott, and Windle]{polson2012bayesian}
Nicholas~G Polson, James~G Scott, and Jesse Windle.
\newblock Bayesian inference for logistic models using polya-gamma latent
  variables.
\newblock \emph{arXiv preprint arXiv:1205.0310}, 2012.

\bibitem[Rasmussen and Williams(2006)]{rasmussen2006gaussian}
Carl~Edward Rasmussen and Christopher K.~I. Williams.
\newblock \emph{Gaussian Processes for Machine Learning}.
\newblock MIT Press, 2006.
\newblock ISBN 9780262182539.

\bibitem[Sanders et~al.(2011)Sanders, Bittner, Rosenfeld, Redding, and
  Goss]{sanders2011pulmonary}
Don~B Sanders, Rachel~CL Bittner, Margaret Rosenfeld, Gregory~J Redding, and
  Christopher~H Goss.
\newblock Pulmonary exacerbations are associated with subsequent fev1 decline
  in both adults and children with cystic fibrosis.
\newblock \emph{Pediatric pulmonology}, 46\penalty0 (4):\penalty0 393--400,
  2011.

\bibitem[Shi et~al.(2005)Shi, Murray-Smith, and
  Titterington]{shi2005hierarchical}
Jian~Qing Shi, Roderick Murray-Smith, and DM~Titterington.
\newblock Hierarchical gaussian process mixtures for regression.
\newblock \emph{Statistics and Computing}, 15\penalty0 (1):\penalty0 31--41,
  2005.

\bibitem[Shi et~al.(2007)Shi, Wang, Murray-Smith, and
  Titterington]{shi2007gaussian}
JQ~Shi, B~Wang, Roderick Murray-Smith, and DM~Titterington.
\newblock {Gaussian process functional regression modeling for batch data}.
\newblock \emph{Biometrics}, 63\penalty0 (3):\penalty0 714--723, 2007.

\bibitem[Song et~al.(2002)Song, Davidian, and Tsiatis]{song2002semiparametric}
Xiao Song, Marie Davidian, and Anastasios~A Tsiatis.
\newblock A semiparametric likelihood approach to joint modeling of
  longitudinal and time-to-event data.
\newblock \emph{Biometrics}, 58\penalty0 (4):\penalty0 742--753, 2002.

\bibitem[Szczesniak et~al.(2013)Szczesniak, McPhail, Duan, Macaluso, Amin, and
  Clancy]{szczesniak2013semiparametric}
Rhonda~D Szczesniak, Gary~L McPhail, Leo~L Duan, Maurizio Macaluso, Raouf~S
  Amin, and John~P Clancy.
\newblock A semiparametric approach to estimate rapid lung function decline in
  cystic fibrosis.
\newblock \emph{Annals of epidemiology}, 23\penalty0 (12):\penalty0 771--777,
  2013.

\bibitem[Taylor et~al.(2013)Taylor, Park, Ankerst, Proust-Lima, Williams,
  Kestin, Bae, Pickles, and Sandler]{taylor2013real}
Jeremy~MG Taylor, Yongseok Park, Donna~P Ankerst, Cecile Proust-Lima, Scott
  Williams, Larry Kestin, Kyoungwha Bae, Tom Pickles, and Howard Sandler.
\newblock Real-time individual predictions of prostate cancer recurrence using
  joint models.
\newblock \emph{Biometrics}, 69:\penalty0 206--213, 2013.

\bibitem[VandenBranden et~al.(2012)VandenBranden, McMullen, Schechter, Pasta,
  Michaelis, Konstan, Wagener, Morgan, and McColley]{vandenbranden2012lung}
Stacy~L VandenBranden, Ann McMullen, Michael~S Schechter, David~J Pasta, Rory~L
  Michaelis, Michael~W Konstan, Jeffrey~S Wagener, Wayne~J Morgan, and
  Susanna~A McColley.
\newblock Lung function decline from adolescence to young adulthood in cystic
  fibrosis.
\newblock \emph{Pediatric pulmonology}, 47\penalty0 (2):\penalty0 135--143,
  2012.

\bibitem[Vonesh et~al.(2006)Vonesh, Greene, and Schluchter]{vonesh2006shared}
Edward~F Vonesh, Tom Greene, and Mark~D Schluchter.
\newblock Shared parameter models for the joint analysis of longitudinal data
  and event times.
\newblock \emph{Statistics in medicine}, 25\penalty0 (1):\penalty0 143--163,
  2006.

\bibitem[Wang et~al.(2010)Wang, Dey, et~al.]{wang2010generalized}
Xia Wang, Dipak~K Dey, et~al.
\newblock Generalized extreme value regression for binary response data: an
  application to b2b electronic payments system adoption.
\newblock \emph{The Annals of Applied Statistics}, 4\penalty0 (4):\penalty0
  2000--2023, 2010.

\bibitem[Zellner(1986)]{zellner1986assessing}
Arnold Zellner.
\newblock On assessing prior distributions and bayesian regression analysis
  with g-prior distributions.
\newblock \emph{Bayesian inference and decision techniques: Essays in Honor of
  Bruno De Finetti}, 6:\penalty0 233--243, 1986.

\end{thebibliography}

\end{document}